\documentclass[twocolumn]{aastex63}

\usepackage{natbib}
\usepackage{color}
\usepackage{mathtools}
\usepackage{hyperref} 

\def    \apjl  		{\rm {ApJL}}

\def    \apjl  		{\rm {ApJL}}

\def	\cm		{\,{\rm {cm}}}
\def	\K		{\,{\rm K}}
\def	\g		{\,{\rm {g}}}
\def	\mum	{\,{\mu \rm{m}}}

\def \bea {\begin{eqnarray}}
\def \ena {\end{eqnarray}}

\def    \bB     {\bf  B}

\def	\bB	{{\bf B}}
\def	\B	{{\rm B}}

\def	\bJ	{{\bf J}}
\def	\bk	{{\bf k}}

\def    \bmu    {{\hbox{\boldsym\char'026}}}	

\def	\C	{{\rm C}}

\def	\cm	{\,{\rm cm}}
\def	\km	{\,{\rm km}}

\def	\max	{\,{\rm max}}
\def	\d	{{\rm d}}

\def	\D	{{\rm D}}
\def	\eff	{{\rm eff}}

\def	\g	{\,{\rm g}}
\def	\gas	{\,{\rm gas}}
\def	\G	{{\rm G}}

\def	\H	{{\rm H}}

\def	\N	{{\rm N}}

\def	\s	{\,{\rm s}}
\def	\sp	{{\rm sp}}

\def	\V	{{\rm V}}
\def	\Bar	{{\rm Bar}}
\def	\rad	{{\rm rad}}
\def	\yr	{{\rm yr}}

\def	\xhat		{\hat{\bf x}}

\def	\ahat		{\hat{\bf a}}

\def    \Bv     	{\bf  B}

\def    \gas     	{{\rm gas}}
\def    \erf     	{{\rm erf}}

\font\mib=cmmib10

\def\bOmega{\hbox{\mib\char"0A}}
\def\bmu{\hbox{\mib\char"16}}

\begin{document}
\shorttitle{Effect of magnetic fields on grain growth}
\shortauthors{Thiem Hoang}
\title{Effects of grain alignment with magnetic fields on grain growth and the structure of dust aggregates}

\author{Thiem Hoang}
\affiliation{Korea Astronomy and Space Science Institute, Daejeon 34055, Republic of Korea, \href{mailto:thiemhoang@kasi.re.kr}{thiemhoang@kasi.re.kr}}
\affiliation{Korea University of Science and Technology, 217 Gajeong-ro, Yuseong-gu, Daejeon, 34113, Republic of Korea}

\begin{abstract}
Dust grains drift through the interstellar medium (ISM) and are aligned with the magnetic field. Here we study the effect of grain alignment and grain motion on grain growth in molecular clouds (MCs). We first discuss the characteristic timescales of alignment of the grain axis of maximum inertia ($\ahat_{1}$) with its angular momentum (${\bf J}$) (i.e., {\it internal alignment}) and alignment of ${\bf J}$ with the magnetic field ($\Bv$, i.e., {\it external alignment}). We determine the maximum grain size with efficient internal ($a_{\max,aJ}$) and external alignment ($a_{\max,JB}$) for composite grains. For the MC density of $n_{\H}\sim 10^{3}-10^{8}\cm^{-3}$, we find that external alignment can occur for very large grains, but internal alignment only occurs for grains smaller than $a_{\max,aJ}\sim 2\mum$. The presence of iron clusters within dust grains or suprathermal rotation increase $a_{\max,aJ}$ to $\sim 10-50\mum$. We then study the growth of aligned grains drifting through the gas. Due to the motion of aligned grains across the magnetic field, gas accretion would increase the grain elongation rather than decrease as expected from the growth of randomly oriented grains. Coagulation by grain collisions also increases grain elongation, leading to the increase of elongation with the grain size. The coagulation of aligned grains forms dust aggregates that contain elongated binaries comprising a pair of grains with parallel short axes. Grains within dust aggregates in 67P/Churyumov-Gerasimenko obtained by {\it Rosetta} have the grain elongation increasing with the grain radius, implying that such dust aggregates might form from aligned grains.

\end{abstract}

\section{Introduction}
Dust is an essential component of the interstellar medium (ISM) and plays an important role in numerous astrophysical processes, including star and planet formation, gas heating and cooling, and formation of water and complex molecules. Dust is mainly formed in the envelope of evolved stars and ejecta of core-collapse supernovae. Upon its injection into the ISM, dust is reprocessed due to gas-grain and grain-grain collisions. Grain growth by gas accretion and grain coagulation are expected to efficiently occur in MCs. The formation of an ice mantle on the grain core at the visual extinction of $A_{V}\gtrsim 3$ (\citealt{1983Natur.303..218W}) is frequently suggested to facilitate grain growth. For instance, \cite{Ormel:2009p4022} suggested that dust coagulation can form micron-sized grains in MCs if the grains are coated with an ice mantle such that the sticking coefficient is enhanced, allowing grain growth upto the shattering threshold of $v_{\rm shat}\sim 2.7\km\s^{-1}$ (see also \citealt{2013MNRAS.434L..70H}). 

The efficiency of grain growth essentially depends on the dust physical properties and grain motion. Grain motion relative to the gas (e.g., drift) is common in the ISM due to various acceleration mechanisms, including radiation pressure, magnetohydrodynamic (MHD) turbulence, shocks. For instance, charged dust grains can be accelerated by MHD turbulence through the gyroresonance effect (\citealt{2003ApJ...592L..33Y}; \citealt{Yan:2004ko}; \citealt{Hoang:2012cx}) or transit time damping \citep{Hoang:2012cx}. The grain motion due to gyroresonance acceleration is previously found to be important for grain coagulation and shattering by grain-grain collisions (\citealt{Hirashita:2009p1139}; \citealt{Hirashita.2021pqi}). 

Previous studies on grain growth usually ignore the effect of magnetic fields and assume spherical grains or randomly oriented grains. Nevertheless, dust grains in general have non-spherical shapes (e.g., elongated shape) and are systematically aligned with the magnetic field, as revealed by starlight polarization (\citealt{Hall:1949p5890}; \citealt{Hiltner:1949p5851}) and polarization of thermal dust emission (e.g., \citealt{PlanckCollaboration:2015ev}). Moreover, modern theory based on Radiative Torques (RATs; \citealt{2007MNRAS.378..910L}; \citealt{Hoang:2008gb}) implies that grain alignment with the magnetic field is efficient in most astrophysical environments, from the ISM to dense MCs (\citealt{2014MNRAS.438..680H}; \citealt{Hoang.2021}) where grain growth can occur efficiently. Indeed, multiwavelength polarization observations reveal that grains are still aligned at high visual extinction of $A_{V}> 20$ within dense clouds (\citealt{2017ApJ...849..157W}; \citealt{Vaillancourt:2020ch}). Therefore, grain growth naturally involves aligned grains instead of randomly oriented grains. Because grains tend to align with their shortest axes perpendicular to the local magnetic field (\citealt{2015ARA&A..53..501A}; \citealt{LAH15}), grain growth from collisions between aligned grains would be dramatically different from randomly oriented grains and would leave fundamental features in the large composite particles resulting from grain collisions.

Another interesting property of grain motion in the magnetized turbulent ISM is that grain acceleration by MHD turbulence is mostly perpendicular to the magnetic field because gyroresonance tends to increase the pitch angle between the grain motion and the magnetic field (\citealt{2003ApJ...592L..33Y}; \citealt{Yan:2004ko}). 
In star-forming clouds, ambipolar diffusion also induces grain motion perpendicular to the ambient magnetic field. The direction of the grain's motion is therefore parallel to the grain long axis, and both are perpendicular to the magnetic field. This fundamental property may affect the shape and internal structure of dust aggregates formed by gas accretion and grain-grain collisions because the collisional cross-section becomes anisotropic. The main goal of this paper is to study the effect of grain alignment and grain motion with respect to the magnetic field on the grain shape and structure, and discuss the implications for observational constraints of dust physics.

The paper is organized as follows. In Section \ref{sec:physics}, we review the leading processes that induce the grain motion in perpendicular to the magnetic field. In Section \ref{sec:alignment}, we discuss characteristic timescales relevant in grain alignment and identify the range of grain sizes that have efficient alignment with the magnetic field for the different grain magnetic properties. In Sections \ref{sec:accretion} and \ref{sec:growth}, we study the grain growth due to gas accretion and grain-grain collisions for aligned grains, respectively. Discussion and summary of our main results are presented in Section \ref{sec:discuss} and \ref{sec:summary}, respectively.

\section{Grain motion perpendicular to the magnetic field}\label{sec:physics}
We first discuss the main processes inducing grain motion in the direction perpendicular to the magnetic field, including gyroresonance acceleration, ambipolar diffusion, and shocks (see also \citealt{Lazarian.2020bwy}).

\subsection{Gyroresonance acceleration by MHD turbulence}
\begin{figure}
\includegraphics[width=0.5\textwidth]{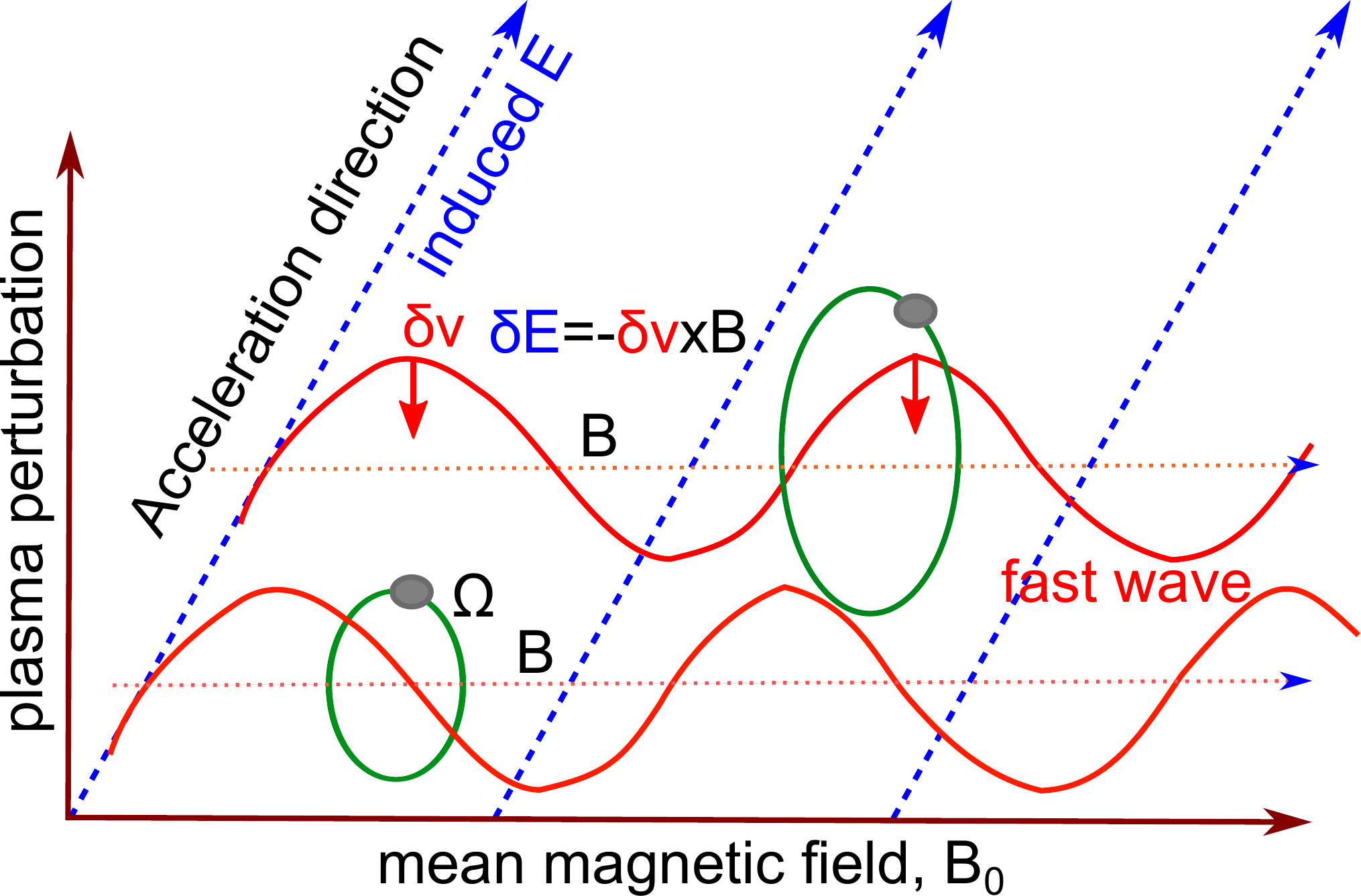}
\caption{Illustration of grain acceleration by gyroresonance. A charged grain gyrates around the mean field (green loop). Fluctuations of fluid induced by fast modes induces an electric field $\delta {\bf E}$ perpendicular to the mean field, which accelerates charged grains in this direction.}
\label{fig:gyro}
\end{figure}

MHD turbulence is found to efficiently accelerate charged grains (\citealt{Yan:2004ko}; \citealt{Hoang:2012cx}). Figure \ref{fig:gyro} shows the schematic illustration of grain acceleration by gyroresonance, leading to the grain motion perpendicular to the mean magnetic field. The velocity of grains due to MHD acceleration is found to change with environments (\citealt{Hoang:2012cx}).

For both silicate and carbonaceous grains in MCs, the grain velocity is approximately equal to $v_{d}\sim 0.4-1\km\s^{-1}$ for grain radius $>0.05\mum$, and $v_{d}$ decreases rapidly for smaller grains. For dense clouds (DC), the grain velocity is smaller of $v_{d}\sim 0.3-1 \km\s^{-1}$. In the Cold Neutral Medium (CNM), the grain velocity can reach $v_{d}\sim 1\km\s^{-1}$ \citep{Yan:2004ko}. Note that, for a gas of temperature $T_{\rm gas}$, the thermal velocity of hydrogen atoms of mass $m_{\rm H}$ is $v_{\rm T}=\left(2k_{\B}T_{\rm gas}/m_{\rm H}\right)^{1/2}\approx 0.4T_{1}\km\s^{-1}$ where $T_{1}=T_{\gas}/10\K$. Thus, MHD turbulence can induce subsonic to supersonic motion of grains across the interstellar magnetic field, depending on the grain size and the gas properties.

\subsection{Ambipolar diffusion}
In star-forming magnetized MCs, such as dense cores or prestellar cores, charged grains are tied to the magnetic field but neutral gas can cross the field and experiences gravitational collapse to form a protostar. The gas-dust collisions will then drag dust together with the neutral gas. Such an ambipolar diffusion effect results in the drift of grains in the direction perpendicular to the magnetic field. Gas accretion is then anisotropic similar to gyroresonance acceleration. The ambipolar diffusion velocity is $v_{d}\sim 0.2-0.3\km\s^{-1}$ for grains of size $a\sim 0.01-0.1\mum$ (\citealt{1995ApJ...453..238R}). Thus, ambipolar diffusion can induce subsonic drift of charged grains relative to the neutral gas. 

\subsection{Shocks in magnetized media}
Shocks are present in dense, magnetized MCs due to young stellar outflows and supernovae. The shock can induce the grain motion perpendicular to the magnetic field. The idea is that charged grains gyrate around the magnetic field and tied to the field. Neutral gas-dust drag can then induce the grain motion across the magnetic field. The drift velocity of dust relative to the neutral gas depends on the shock speed and the magnetic field strength and can reach supersonic motion \citep{2019ApJ...877...36H}.

Note that the efficiency of the aforementioned acceleration mechanisms depend on the grain charge. In general, grain charging is induced by collisions with electrons and ions in the gas, photoelectric effect by UV photons, and ionization by cosmic rays (e.g., \citealt{Hoang:2012cx}).

\section{Grain alignment with magnetic fields}\label{sec:alignment}
The alignment of an irregular grain with the magnetic field in general can be distinguished into two stages: (i) the alignment of the grain axis of maximum inertia moment (shortest axis) parallel to its angular momentum, ${\bf J}$ (hereafter internal alignment), and (ii) the alignment of ${\bf J}$ with the magnetic field (hereafter external alignment; see \citealt{2007JQSRT.106..225L}; \citealt{LAH15} for recent reviews). When the internal alignment occurs with the grain's shortest axis perpendicular to ${\bf J}$, it is called {\it wrong internal alignment}. In this section, we review the main processes involved in grain alignment in MCs and show that submicron-sized grains of ordinary paramagnetic material or micron-sized grains with iron inclusions can be efficiently aligned with the grain shortest axis parallel to the magnetic field.

\subsection{Grain model and assumptions}
A grain of irregular shape is described by the principal axes $\ahat_{1},\ahat_{2},\ahat_{3}$. Let $I_{1}> I_{2}\ge I_{3}$ be the principal moments of inertia along the principal axes, respectively. For convenience of numerical estimates, throughout this paper, we assume an oblate spheroidal grain, such that the principal moments of inertia along the principal axes are $I_{1}> I_{2}=I_{3}$. Let us denote $I_{\|}\equiv I_{1}$ and $I_{\perp}=I_{2}$ for simplicity. The length of the symmetry (semi-minor) axis is denoted by $c$ and the lengths of the semi-major axes are denoted by $a$ and $b$ with $a=b$ (see Figure \ref{fig:torque-free}). 

For a general case, the grain angular momentum ($\bJ$) and angular velocity ($\bOmega$) are not parallel to the axis of maximum inertia, $\ahat_{1}$. So let $\theta$ be the angle between $\bJ$ and $\ahat_{1}$. The component of the angular velocity projected onto the $\ahat_{1}$ is $\Omega_{1}=J_{1}/I_{1}=J\cos\theta/I_{1}\equiv \Omega_{0}\cos\theta$ with $\Omega_{0}=J/I_{1}$ (see Figure \ref{fig:torque-free}).

The principal moments of inertia for the rotation parallel and perpendicular to the grain symmetry axis are given by
\bea
I_{\|}&=&\frac{8\pi}{15}\rho a^{4}c=\frac{8\pi}{15}\rho sa^{5},\label{eq:Ipar}\\
I_{\perp}&=&\frac{4\pi}{15}\rho a^{2}c\left(a^{2}+c^{2}\right)=\frac{4\pi}{15}\rho sa^{5}\left(1+s^{2}\right),\label{eq:Iparperp}
\ena 
where $\rho$ is the mass density of the grain, and $s=c/a<1$ is the axial ratio. The ratio of the principal inertia moments is $h=I_{\|}/I_{\perp}=2/(1+s^{2})>1$ for $s<1$ (see also \citealt{2014MNRAS.438..680H}). The effective grain size $a_{\eff}$ is defined as the equivalent sphere of the same volume, so $a_{\eff}=(ca^{2})^{1/3}=s^{1/3}a$.

In this paper, we consider a composite grain model whose constituents include primarily amorphous silicate, Fe clusters (metallic inclusions), and amorphous carbon. The composite grain model is the natural outcome of grain coagulation in MCs under our interest, and also aligned with the "Astrodust" model recently proposed by \cite{Draine.2021b1e}.

\subsection{Grain rotational damping}

The grain rotation can be damped by various processes, including sticky collisions of gas species followed by evaporation and infrared emission (see e.g., \citealt{Hoang.2021}). The rotational damping time due to gas collisions is given by
\bea
\tau_{\gas}&=&\frac{3}{4\sqrt{\pi}}\frac{I_{\|}}{1.2n_{\rm H}m_{\rm H}
v_{\rm T}a^{4}\Gamma_{\|}}\nonumber\\
&\simeq& 8.3\times 10^{3}\hat{\rho}sa_{-5}\left(\frac{1}{n_{3}T_{1}^{1/2}\Gamma_{\|}}\right)~{\rm yr},\label{eq:tgas}
\ena
where $a_{-5}=a/(10^{-5}\cm)$, $\hat{\rho}=\rho/(3\g\cm^{-3})$, and $\Gamma_{\|}$ is the geometrical factor of unity order (\citealt{1993ApJ...418..287R}; \citealt{2009ApJ...695.1457H}). The gas density is normalized to its typical value of a dense MC with $n_{3}=n_{\H}/(10^{3}\cm^{-3})$. For dense MCs, the damping by infrared emission is subdominant and disregarded in this study for simplicity.

In addition to rotational damping, grains can also be spun-up due to evaporation of molecules from the grain surface. In thermal equilibrium, the grain rotational kinetic energy is equal to the gas thermal energy. We can then define the thermal angular velocity as $\Omega_{T}=(kT_{\rm gas}/I_{\|})^{1/2}\simeq 7.4\times 10^{4}\hat{\rho}\hat{s}a_{-5}^{-5/2}T_{1}^{1/2}\rad\s^{-1}$ and the thermal angular momentum $J_{T}=I_{\|}\Omega_{T}=(I_{\|}kT_{\gas})^{1/2}$.

\subsection{Grain magnetic properties}
In order to interact and align with the ambient the magnetic field, dust grains must have magnetic susceptibility. Amorphous silicate grains or dust grains with iron inclusions are paramagnetic (PM) material due to the existence of unpaired electrons. Atomic nuclei within the grain can also have {\it nuclear} paramagnetism due to unpaired protons and nucleons. We first describe the magnetic susceptibility in this section.

\subsubsection{Magnetic susceptibility from electron spins}
An atom with unpaired electrons with the angular momentum $\bJ$ has the magnetic moment $\bmu_{p}=-g_{e}e/(2m_{e}c){\bf J}=\gamma_{e}\bJ$ with $e$ the elementary charge, $g_{e}\approx 2$ the electron $g$-factor, and $\gamma_{e}=-g_{e}e/(2m_{e}c)$ is the gyromagnetic ratio of electron spins. 

The magnitude of the atomic magnetic moment is then given by
\bea
\mu_{p}=g_{e}\mu_{B}\sqrt{J(J+1)}\equiv p\mu_{\B},\label{eq:mu}
\ena
where $\mu_{B}=e\hbar/2m_{e}c$ the Bohr magneton,
and $p=g_{e} \sqrt{J(J+1)}$ with $J$ the angular momentum quantum number of electrons in the outer partially filled shell (see \citealt{Draine:1996p6977} and references therein).

Let $f_{p}$ is the fraction of atoms which are paramagnetic (e.g., Fe) in the silicate component. The number density of paramagnetic atoms in the dust grain is then $n_{p}=f_{p}n$ with $n$ the total number density of atoms.

The zero-frequency susceptibility $\chi(0)$ of a paramagnetic material at dust temperature $T_{d}$ is described by the Curie's law:
\bea
\chi(0)&=&\frac{n_{p}\mu_{p}^{2}}{3k_{\B}T_{\d}},\label{eq:curielaw}
\ena
which corresponds to
\bea
\chi(0)\simeq 0.06f_{p}n_{23}\hat{p}^{2}\left(\frac{10\K}{T_{d}}\right),\label{eq:chi_para}
\ena
where $n_{23}=n/10^{23}\cm^{-3}$ and $\hat{p}=p/5.5$. For silicate of MgFeSiO$_{4}$ structure which accommodates $100\%$ of Fe abundance since $X(Si)\approx X(Fe)$, one has $f_{p}=1/7$ and $J=5/2$ (see \citealt{2016ApJ...831..159H}). However, a lower fraction of Fe in silicate is expected, and the value of $f_{p}$ is then lower. Here, we take $f_{p}=0.01$ for a typical value.\footnote{For the amorphous silicate of structure $Mg_{1.3}(Fe,Ni)_{0.3}SiO_{3.6}$ which accommodates $30\%$ of Fe in the Astrodust model \citep{Draine.2021b1e}, $f_{p}=0.3/(1.3+0.3+1+3.6)\approx 0.05$. The mass density of amorphous silicate is $\rho_{sil}=3.41\g/\cm^{3}$, so $n=3.41/\bar{m}= 9.5\times 10^{22}\approx 10^{23}\cm^{-3}$ where the mean mass per atom $\bar{m}=(1.3m_{Mg}+0.3m_{Fe}+m_{Si}+3.6m_{O})/6.2$.}

The presence of iron clusters within the composite grain makes the dust become superparamagnetic material because the iron clusters act as giant magnetic moments. Let $N_{\rm cl}$ be the number of Fe atoms per cluster and $\phi_{\rm sp}$ be the volume filling factor of iron clusters in the composite grain. The zero-frequency magnetic susceptibility is enhanced significantly to (\citealt{2016ApJ...831..159H}):
\bea
\chi_{\rm sp}(0)\approx 0.052N_{\rm cl}\phi_{\rm sp}\hat{p}^{2}\left(\frac{10\K}{T_{d}}\right),\label{eq:chi_sp}
\ena
which is about a factor $N_{\rm cl}$ greater than $\chi(0)$ of ordinary paramagnetic material (Eq. \ref{eq:chi_para}). The possible value of $N_{\rm cl}$ spans from $\sim 20$ to $10^{5}$ (\citealt{Jones:1967p2924}), and $\phi_{\rm sp}\sim 0.3$ if $100\% $ of Fe abundance present in iron clusters (see e.g., \citealt{2016ApJ...831..159H}). For our following calculations, we consider two values of $\phi_{\rm sp}=0$ (i.e., ordinary paramagnetic grains) and $\phi_{\rm sp}=0.03$ (e.g. \citealt{1999ApJ...512..740D}) for superparamagnetic grains.



The imaginary part of the complex magnetic susceptibility is a function of frequency and usually represented as $\chi_{2}(\omega)=\omega K(\omega)$ where $K(\omega)$ is the function obtained from solving the magnetization dynamics equation. 
 
For paramagnetic material, $K(\omega)$ can be described by the critically-damped solution (\citealt{1999ApJ...512..740D})
\bea
K(\omega)& =&\frac{\chi(0)\tau_{2}}{[1+(\omega \tau_{2}/2)^{2}]^{2}}\nonumber\\
&\simeq& 1.7\times 10^{-13}\hat{p}\left(\frac{10\K}{T_{d}}\right)\frac{1}{[1+(\omega \tau_{2}/2)^{2}]^{2}}\s,\label{eq:kappa_para}
\ena
where $\tau_{2}\approx 2.9\times 10^{-12}/f_{p}$ s is the electron spin-spin relaxation time. The susceptibility $K(\omega)$ is constant for low frequency but decreases rapidly at high frequency of $\omega>2/\tau_{2}\sim 10^{12}f_{p} \rad\s^{-1}$. So, for large grains with lower rotation frequency, $K$ is independent of $\omega$.

Similarly, for superparamagnetic material, $K_{\rm sp}(\omega)$ is given by (see \citealt{2016ApJ...831..159H}):
\bea
K_{\sp}(\omega) &=& \frac{\chi_{\sp}(0)\tau_{\sp}}{[1+(\omega \tau_{\sp}/2)^{2}]^{2}},\nonumber\\
&\simeq & 2.6\times 10^{-11}N_{\rm cl}\phi_{\rm sp}\hat{p}^{2}\frac{\exp\left(N_{\rm cl}T_{\rm act}/T_{d}\right)}{\hat{T}_{d}[1+(\omega \tau_{\sp}/2)^{2}]^{2}}\s,~~~\label{eq:kappa_sp}
\ena
where $\tau_{\rm sp}$ is the timescale of thermally activated remagnetization given by
\bea
\tau_{\sp}\approx \nu_{0}^{-1} \exp\left(\frac{N_{\rm cl}T_{\rm act}}{T_{d}}\right)
\label{eq:tau_sp}
\ena
where experiments give $\nu_0\approx 10^{9}\s^{-1}$ and $T_{\rm act}\approx 0.011\K$ (see \citealt{Morrish:2001vp}). The susceptibility $K_{\rm sp}(\omega)$ is constant for low frequency but decreases rapidly at high frequency for $\omega>2/\tau_{\rm sp}\sim 2\nu_{0}\sim 10^{9}\rad\s^{-1}$.

\subsubsection{Magnetic susceptibility from nuclear spins}
A nucleus with unpaired nucleons can also induce magnetism due to the intrinsic magnetic moment of protons and neutrons (see e.g., \citealt{1999ApJ...520L..67L}, hereafter LD99).


Let $\mu_{n}$ be the magnetic moment of a nucleus in the dust grain. The zero-frequency magnetic susceptibility of the grain can also be described by the Curie's law:
\bea
\chi_{n}(0)&=&\left(\frac{n_{n}\mu_{n}^{2}}{3kT_{d}}\right),\nonumber\\
&\approx&4.5\times 10^{-10}\left(\frac{\mu_{n}}{2.79\mu_{N}}\right)^{2}\left(\frac{n_{n}}{10^{22}\cm^{-3}}\right)\left(\frac{10\K}{T_{d}}\right),\label{eq:chi_n}
\ena
where $n_{n}$ is the number density of nuclei in the composite grain that has the magnetic moment $\mu_{n}$, and $\mu_{N}=e\hbar/2m_{p}c=5.05 \times10^{-24}$ erg G$^{-1}$ with $m_{p}$ the proton mass is the nuclear magneton.

As seen in LD99, the hydrogen nucleus (proton) has the largest magnetic moment of $\mu_{n}=2.79\mu_{N}$, whereas heavy elements like Si, Mg and Fe have a much lower magnetic moment. Therefore, as in LD99, we consider the nuclear magnetism induced by hydrogen nuclei only and take $n_{n}=10^{22}\cm^{-3}$ (i.e., the hydrogen fraction is $10\%$ of the dust) as an example. 

The frequency-dependence nuclear susceptibility is given by
\bea
K_{n}(\omega) &=& \frac{\chi_{n}(0)\tau_{n}}{[1+(\omega \tau_{n}/2)^{2}]^{2}}\nonumber\\
&\simeq& 4.5\times 10^{-14}\left(\frac{\chi_{n}(0)}{4.5\times 10^{-10}}\right)\left(\frac{\tau_{n}}{10^{-4}\s}\right)\nonumber\\
&&\times\frac{1}{[1+(\omega \tau_{n}/2)^{2}]^{2}}\s,\label{eq:Kn}
\ena
where $\tau_{n}$ is the time of nuclear spin-spin relaxation which is given by
\bea
\tau_{n}^{-1}=\tau_{ne}^{-1}+\tau_{nn}^{-1},
\ena
where $\tau_{ne}\approx g_{e}\hbar/(3.8n_{e}\mu_{n}^{2})\simeq 2.3\times 10^{-4}(3.1/g_{n})^{2}(10^{22}\cm^{-3}/n_{e})\s$ with $n_e$ the number density of unpaired electrons in the dust and $g_{n}=\mu_{n}/\mu_{N}$, and $\tau_{nn}=\hbar/(3.8n_{n}g_{n}\mu_{n}^{2})\approx 0.16\tau_{ne} (n_{e}/n_{n})$ (see LD99 for more details).

Similar to electron susceptibility, the nuclear susceptibility $K_{n}(\omega)$ is constant for low frequency but decreases rapidly at frequency $\omega>1/\tau_{n}\sim 10^{4}\rad \s^{-1}$. However, nuclear magnetism is suppressed at a much lower frequency than the case of electron magnetism (see Eqs. \ref{eq:kappa_para} and \ref{eq:kappa_sp}).


\subsubsection{Grain magnetic moment}
A rotating paramagnetic grain can thus acquires a magnetic moment thanks to the Barnett effect (see \citealt{Barnett:1915p6353}; \citealt{1969mech.book.....L}) and the rotation of its charged body (\citealt{1972MNRAS.158...63M}; \citealt{1976Ap&SS..43..257D}). The Barnett effect, which is shown to be much stronger than the latter, induces a magnetic moment proportional to the grain angular velocity:
\bea
\bmu_{\Bar}=\frac{\chi(0)V}{\gamma_{g}}\bOmega,\label{eq:muBar}
\ena
where $V=4\pi sa^{3}/3$ is the grain volume, $\chi(0)$ is the magnetic susceptibility of the grain at rest (i.e., zero-frequency susceptibility), and the gyromagnetic ratio $\gamma_{g}=\gamma_{e}=-g_{e}\mu_{B}/\hbar$ for electron spins and $\gamma_{g}=\gamma_{n}=g_{n}\mu_{N}/\hbar$ for nuclear spins.

\subsection{Internal Alignment}
Internal alignment of the grain axis of maximum inertia with the angular momentum is caused by the dissipation of the grain rotational energy due to internal relaxation effects, including Barnett relaxation, nuclear relaxation, and inelastic relaxation (\citealt{1979ApJ...231..404P}; \citealt{1999ApJ...520L..67L}). Here, we provide the basic formulae for references.

\subsubsection{Barnett, super-Barnett, and nuclear relaxation}

\begin{figure}
\includegraphics[width=0.45\textwidth]{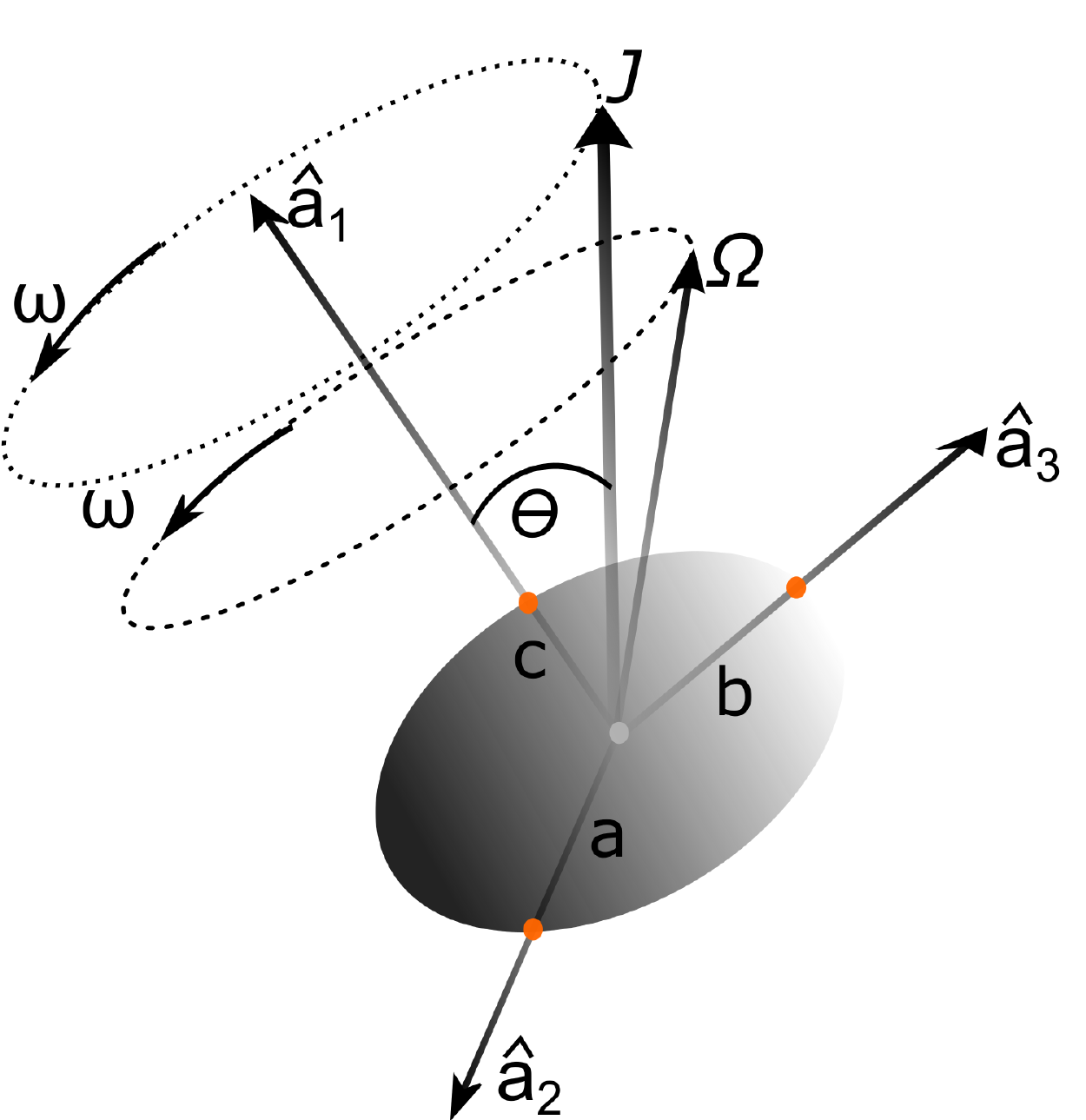}
\caption{Illustration of torque-free motion of an spheroidal grain of principal axes $\ahat_{1}\ahat_{2}\ahat_{3}$. In the body frame, the angular momentum ($\bJ$) and angular velocity ($\bOmega$) are both precessing around the axis of maximum inertia $\ahat_{1}$ with the angular rate $\omega$. Efficient internal relaxation induces the alignment of both $\bJ$ and $\bOmega$ with $\ahat_{1}$ (internal alignment).}
\label{fig:torque-free}
\end{figure}

Consider a grain rotating with $\bJ$ and $\bOmega$ not aligned with the axis of maximum inertia $\ahat_{1}$. In the grain's body frame, the tip of $\bJ$ and $\bOmega$ precesses around $\ahat_{1}$ with the angular rate $\omega=(h-1)\Omega_{1}=(h-1)J\cos\theta/I_{\|}$ (e.g., \citealt{1979ApJ...231..404P}; \citealt{Hoang:2010jy}). The grain instantaneous magnetization by Barnett effect, ${\bf M}=\bmu_{\rm Bar}/V= (\chi(0)/\gamma_{g}) \bOmega$ (see Eq. \ref{eq:muBar}), has a component perpendicular to $\ahat_{1}$, which is rotating with respect to $\ahat_{1}$ at an angular rate $\omega$ (see Figure \ref{fig:torque-free}). The rotating magnetization component has some lag behind the grain material and induces the dissipation of the grain rotational energy, leading to the internal alignment of $\ahat_{1}$ with $\bOmega$ and $\bJ$ (\citealt{1979ApJ...231..404P}). 

The characteristic timescale for Barnett relaxation (BR) is given by (\citealt{1979ApJ...231..404P}; \cite{1993ApJ...418..287R}; see Appendix \ref{apdx:internal} for the derivation)
\bea
\tau_{\rm BR}&=&\frac{\gamma_{e}^{2}I_{\|}^{3}}{VK(\omega)h^{2}(h-1)J^{2}},\nonumber\\
&\simeq& 0.5\hat{\rho}^{2}a_{-5}^{7}f(\hat{s})\left(\frac{J_{d}}{J}\right)^{2}
\left[1+\left(\frac{\omega\tau_{2}}{2}\right)^{2}\right]^{2}\yr,~~~\label{eq:tauBar}
\ena
where $\hat{s}=s/0.5, f(\hat{s})=\hat{s}[(1+\hat{s}^{2})/2]^{2}$, and $J_{d}=\sqrt{I_{\|}k_{\B}T_{d}/(h-1)}$ is the dust thermal angular momentum (see also \citealt{2014MNRAS.438..680H}).

For superparamagnetic grains, the relaxation time by the Barnett effect (so-called super-Barnett relaxation) is significantly reduced by a factor $N_{\rm cl}$:
\bea
\tau_{\rm BR,sp}&=&\frac{\gamma_{e}^{2}I_{\|}^{3}}{VK_{\rm sp}(\omega)h^{2}(h-1)J^{2}}\nonumber\\
&\approx &0.16 \hat{\rho}^{2}f(\hat{s})a_{-5}^{7}\left(\frac{1}{N_{\rm cl}\phi_{\rm sp,-2}}\right)\left(\frac{J_{d}}{J}\right)^{2}\nonumber\\
&&\times \left[1+\left(\frac{\omega\tau_{\rm sp}}{2}\right)^{2}\right]^{2} \exp\left(\frac{N_{\rm cl}T_{\rm act}}{T_{d}}\right) \yr,
\label{eq:tauBar_sup}
\ena
where $\phi_{\rm sp,-2}=\phi_{\rm sp}/0.01$, which is a factor $\sim N_{\rm cl}$ smaller than the Barnett relaxation time for ordinary paramagnetic material (Eq. \ref{eq:tauBar}). 

Nuclear magnetism can also induce internal relaxation as Barnett effect for electron spins (\citealt{1999ApJ...520L..67L}). The nuclear relaxation (NR) time is given by
\bea
\tau_{\rm NR}&=&\frac{\gamma_{n}^{2}I_{\|}^{3}}{VK_{n}(\omega)h^{2}(h-1)J^{2}},\nonumber\\
&\simeq&125\hat{\rho}^{2}a_{-5}^{7}f(\hat{s})\left(\frac{n_e}{n_n}\right) \left(\frac{J_{d}}{J}\right)^{2}
\left(\frac{g_n}{3.1}\right)^{2}\left(\frac{2.79\mu_N}{\mu_n}\right)^{2}\nonumber\\
&&\times \left[1+\left(\frac{\omega\tau_{n}}{2}\right)^{2}\right]^{2}\s,
\label{eq:tau_nucl}
\ena
which increases rapidly with the grain size as $a^{7}$ and angular momentum as $J^{2}$, as the Barnett relaxation (Eq. \ref{eq:tauBar}). Also, the relaxation time rapidly increases with the frequency at $\omega>2/\tau_{n}$.


The total rate of internal relaxation by Barnett, super-Barnett, and nuclear relaxation for a composite grain is given by 
\bea
\tau_{\rm int}^{-1}=\tau_{\rm BR}^{-1}+\tau_{\rm BR,sp}^{-1}+\tau_{\rm NR}^{-1}.\label{eq:tint}
\ena

\begin{figure*}
\includegraphics[width=0.5\textwidth]{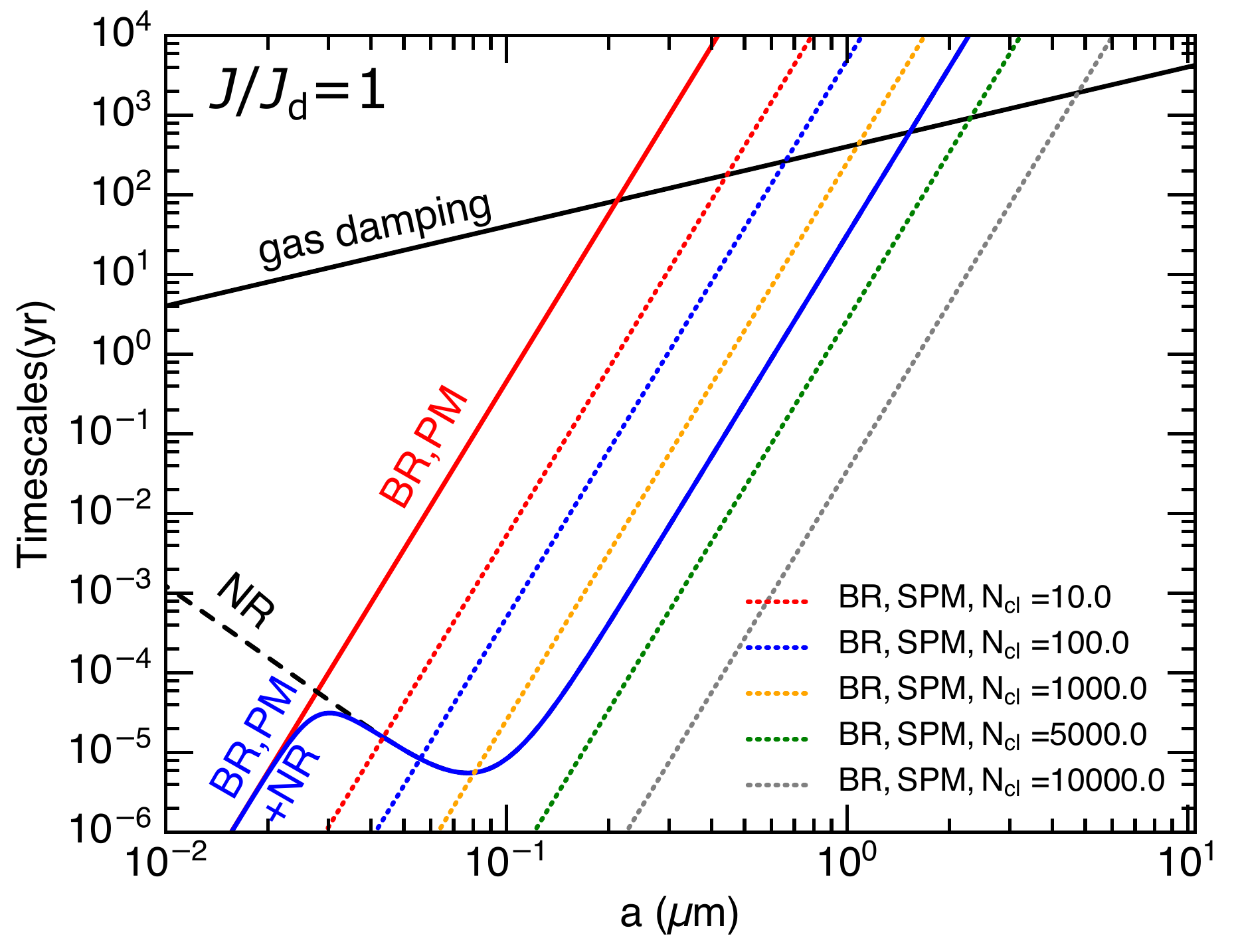}
\includegraphics[width=0.5\textwidth]{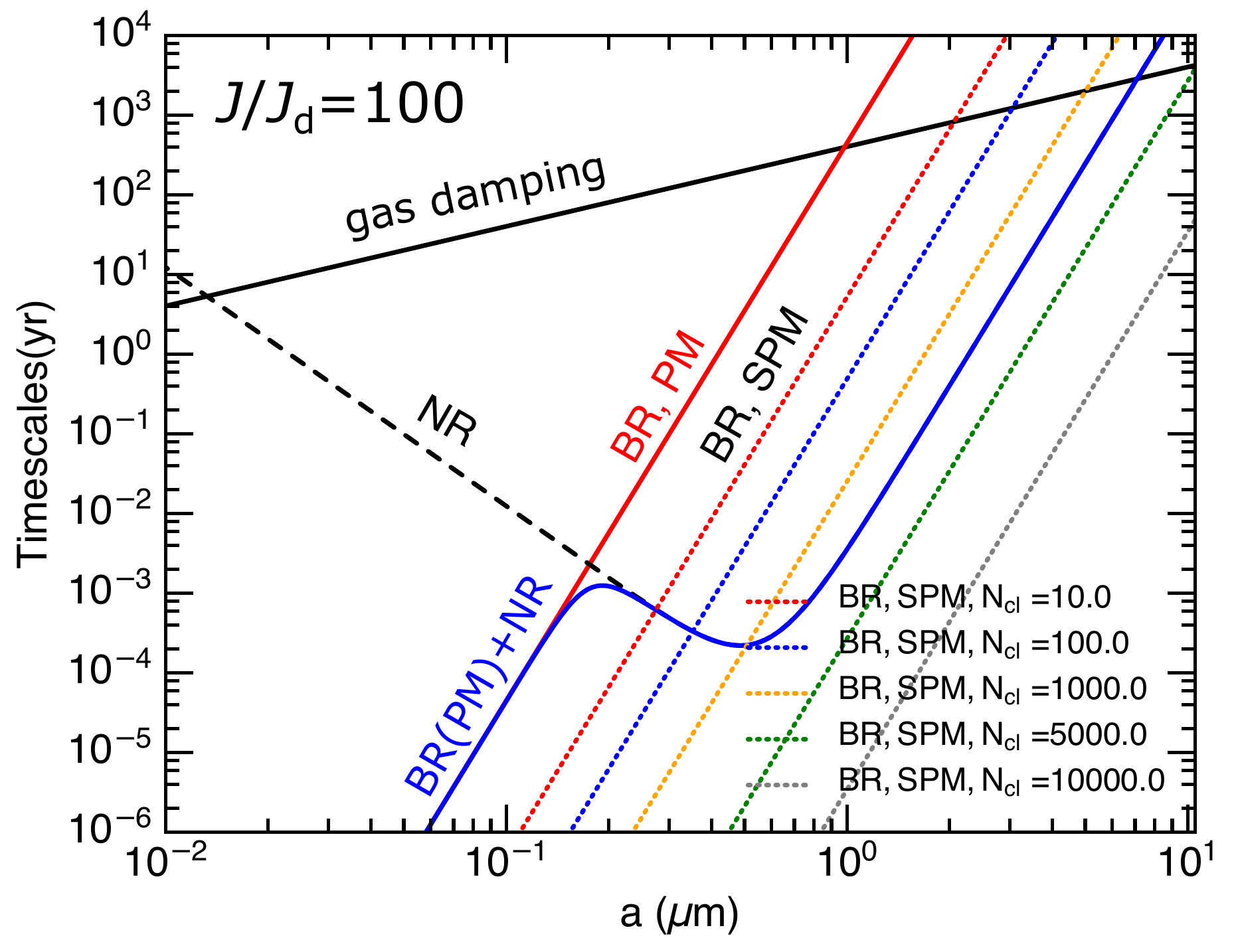}
\caption{Internal relaxation timescales due to Barnett relaxation for paramagnetic (PM) grains (grain without iron clusters), nuclear relaxation, and Barnett relaxation for superparamagnetic (SPM) grains, assuming grain thermal rotation ($J/J_{d}=1$, left panel) and suprathermal rotation ($J/J_{d}=100$, right panel). The gas damping for the DC is also overplotted for comparison. Suprathermal rotation reduces the timescale of Barnett relaxation, but the minimum of nuclear relaxation time is shifted to a larger size, from $a\sim 0.08\mum$ for thermal rotation to $a\sim 0.5\mum$.}
\label{fig:talign_internal}
\end{figure*}

In Figure \ref{fig:talign_internal}, we plot the internal relaxation times for both ordinary paramagnetic grains and superparamagnetic grains with the different level of iron inclusions for two cases of the grain thermal rotation ($J/J_{d}=1$, left panel) and suprathermal rotation ($J/J_{d}=100$, right panel) for the DC condition (see Table \ref{tab:MC} for the chose parameters). The Barnett time increases with the grain size as implied by Equation (\ref{eq:tauBar}), but the nuclear relaxation time increases with the grain size only for large grains of $a>0.1\mum$. For the PM grains of size $a<0.1\mum$, nuclear relaxation changes the trend and $\tau_{NR}$ increases with decreasing the size due to the suppression of the nuclear susceptibility at high angular rate at $\omega>2/\tau_{n}$. Thus, the NR time is much shorter than Barnett relaxation for grains larger than $a\sim 0.03\mum$. The similar trend is observed for suprathermal rotation, but the NR time changes its trend at a larger size of $a\sim 0.05$, and BR is more efficient than nuclear relaxation at $a<0.1\mum$. For SPM grains (dotted lines), the Barnett relaxation time become shorter than nuclear relaxation for $N_{cl}>10^{3}$ or small grain size of $a<0.05\mum$ (left panel) and $a<0.2\mum$ (right panel). In all cases, the internal relaxation time can be longer than the gas damping time for largest grains of $a>1\mum$ and $a>10\mum$ for thermal and suprathermal rotation, respectively.


It is worth noting that \cite{1979ApJ...231..404P} realized that dust grains are not ideal solid and exhibit viscous properties. As a result, the alternating stress caused by the precession of $\bOmega$ with $\ahat_{1}$ lags behind the grain material and induces the dissipation of the grain rotation energy into the heat, resulting in the internal alignment of $\bOmega$ and $\bJ$ with $\ahat_{1}$. However, the characteristic inelastic timescale is uncertain for dust grains due to its unknown modulus of rigidity and elastic $Q$ parameter. Thus, this effect is disregarded in our present paper.

\subsubsection{Maximum size for internal alignment}
\begin{table}[htb]
\begin{center}
\caption{\label{tab:MC}
Physical parameters}
\begin{tabular}{l l l}
\hline\hline
Parameters & MC & DC\cr
\hline
Gas density, $n_{\H}(\cm^{-3})$ & $10^{3}$ & $10^{5}$\cr 
Gas temperature, $T_{\rm gas}(\K)$ & 15 & 15 \cr
Dust temperature, $T_{\rm d}(\K)$ & 15 & 15 \cr
Magnetic field strength, $B (\mu G)$ & 10 & 200 \cr
Fraction of Fe atoms in silicate, $f_{p}$ & 0.01 & 0.01\cr
Volume filling factor of Fe clusters, $\phi_{\rm sp}$ & 0.03 & 0.03\cr
Nuclear proton density, $n_{n}(\cm^{-3})$ & $10^{22}$ & $10^{22}$\cr
Gas Thermal speed, $v_{T}$ & 0.50 & 0.41 \cr
Grain drift velocity, $v_{d}(\km\s^{-1})$ & 0.5 & 0.5\cr
\hline
\cr
\hline\hline
\end{tabular}
\end{center}
\end{table}

As shown in Equations (\ref{eq:tauBar})-(\ref{eq:tau_nucl}),
the relaxation time increases rapidly with the grain size as $a^{7}$. Thus, for sufficiently large grains, the internal relaxation can be slower than the gas randomization (see Figure \ref{fig:talign_internal}), for which the internal alignment ceases. Thus, it is important to estimate the maximum size at which internal relaxation is still efficient to maintain efficient internal alignment. 

We first consider grains with iron inclusions which have the most efficient Barnett relaxation. From Equations (\ref{eq:tgas}) and (\ref{eq:tauBar_sup}), it is straightforward to obtain the ratio of the BR time to the gas damping time:
\bea
\frac{\tau_{\rm BR,sp}}{\tau_{\gas}}&=& \frac{\gamma_{e}^{2}I_{\|}^{2}4\sqrt{\pi}1.2n_{\H}mv_{\rm T}a\Gamma_{\|}}{3Vh^{2}(h-1)J^{2}}\times \frac{[(1+(\omega\tau_{\rm sp}/2)^{2}]^{2}}{\chi_{\rm sp}(0)\tau_{\rm sp}},\nonumber\\
&\approx & 10^{-6}\frac{n_{5}T_{1}^{1/2}a_{-5}^{6}\Gamma_{\|}}{h^{2}N_{\rm cl,4}\phi_{\rm sp,-2}}\left(\frac{J_{d}}{J}\right)^{2}\frac{[(1+(\omega\tau_{\rm sp}/2)^{2}]^{2}}{\exp\left({N_{\rm cl}T_{\rm act}/T_{d}}\right)}.\label{eq:tBR_tgas}
\ena

Equation (\ref{eq:tBR_tgas}) implies the steep increase with the grain size as $a^{6}$, so that sufficiently large grains may not have internal alignment. Let $a_{\max, aJ}$ be the maximum grain size for internal alignment between $\ahat_{1}$ and $\bJ$. The maximum size for internal alignment by Barnett relaxation can be determined by $\tau_{\rm Bar,sp}/\tau_{\gas}=1$, yielding
\bea
a_{\max,aJ}(\rm BR) &=& 1.0h^{1/3} \left(\frac{N_{\rm cl,4}\phi_{\rm sp,-2}}{n_{5}T_{1}^{1/2}\Gamma_{\|}}\right)^{1/6}\left[\exp\left(\frac{N_{\rm cl}T_{\rm act}}{T_{d}}\right)\right]^{1/6}\nonumber\\
&&\times\left(\frac{1}{1+(\omega\tau_{\rm sp}/2)^{2}}\right)^{1/3}\left(\frac{J}{J_{d}}\right)^{1/3}~\mum,\label{eq:aBar}
\ena
which implies $a_{\max,Bar}\approx 1.8\mum$ for $N_{\rm cl}\sim 10^{4}$ and $J/J_{d}=1$ and $n_{5}=1$ (i.e., $n_{\H}=10^{5}\cm^{-3})$. One can see that $a_{\max,aJ}$ increases with iron inclusions as $N_{\rm cl}^{1/6}$, with the rotation rate as $J^{2}$, but decreases with the gas density as $n_{\H}^{-1/6}$.

Similarly, one can find the maximum size for internal alignment induced by nuclear relaxation using Equations (\ref{eq:tgas}) and (\ref{eq:tau_nucl}):
\bea
a_{\max,aJ}(\rm NR) &<& 1.41 h^{1/3}\left(\frac{(n_{e}/n_{n})(g_{n}/3.1)^{2}}{n_{5}T_{1}^{1/2}\Gamma_{\|}}\right)^{1/6}\nonumber\\
&\times&\left(\frac{1}{1+(\omega\tau_{\rm n}/2)^{2}}\right)^{1/3} \left(\frac{J}{J_{d}}\right)^{1/3}~\mum,\label{eq:a_nucl}
\ena
which differs to Equation (\ref{eq:aBar}) by a factor of $[\exp(N_{\rm cl}T_{\rm acc}/T_{d})]^{1/6}/1.4\simeq 4.5$ for $N_{\rm cl}=10^{4}$. The value of $a_{\max,aJ}$ increases with the rotation rate as $J^{2}$, but decreases with the gas density as $n_{\H}^{-1/6}$.

\begin{figure*}
\includegraphics[width=0.5\textwidth]{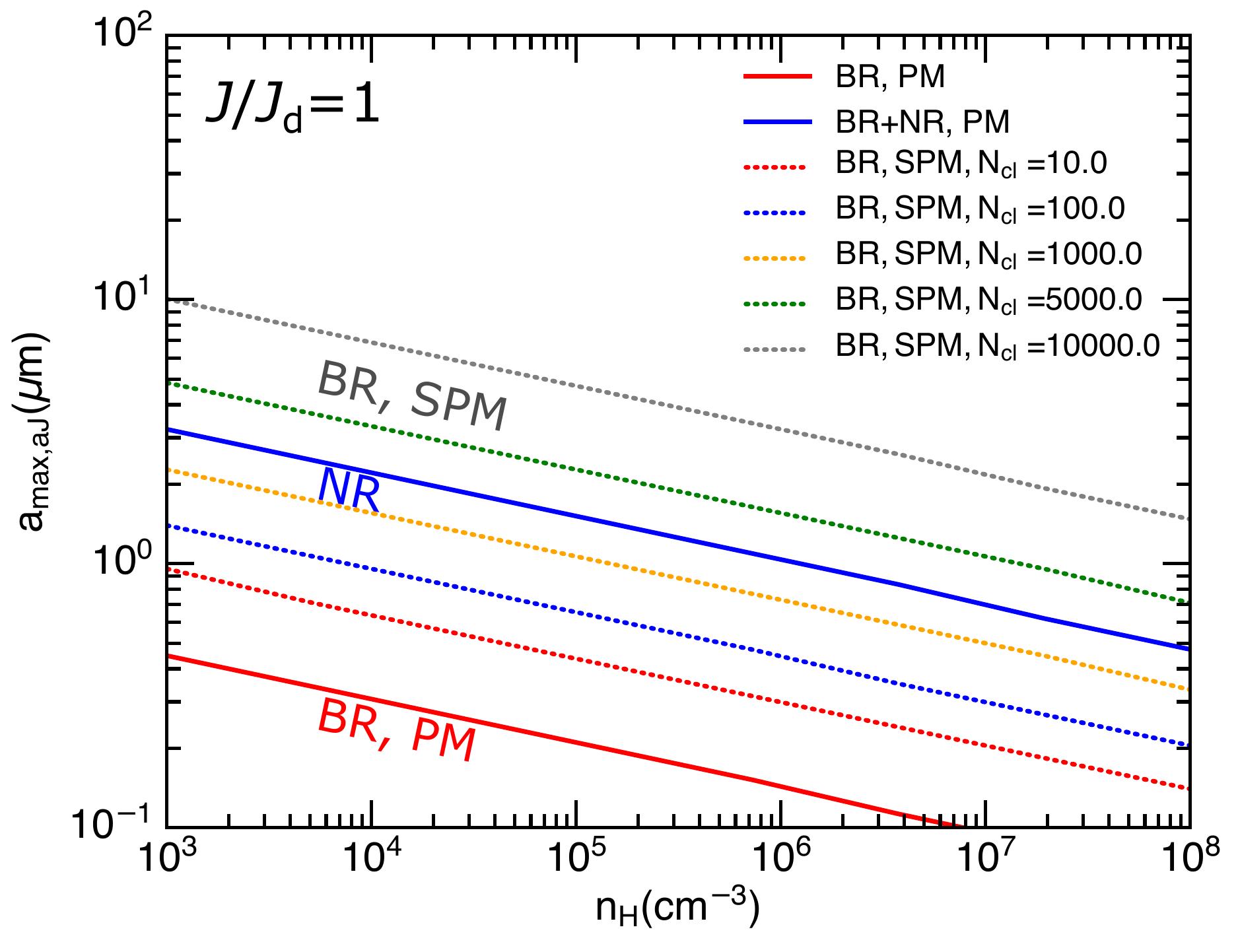}
\includegraphics[width=0.5\textwidth]{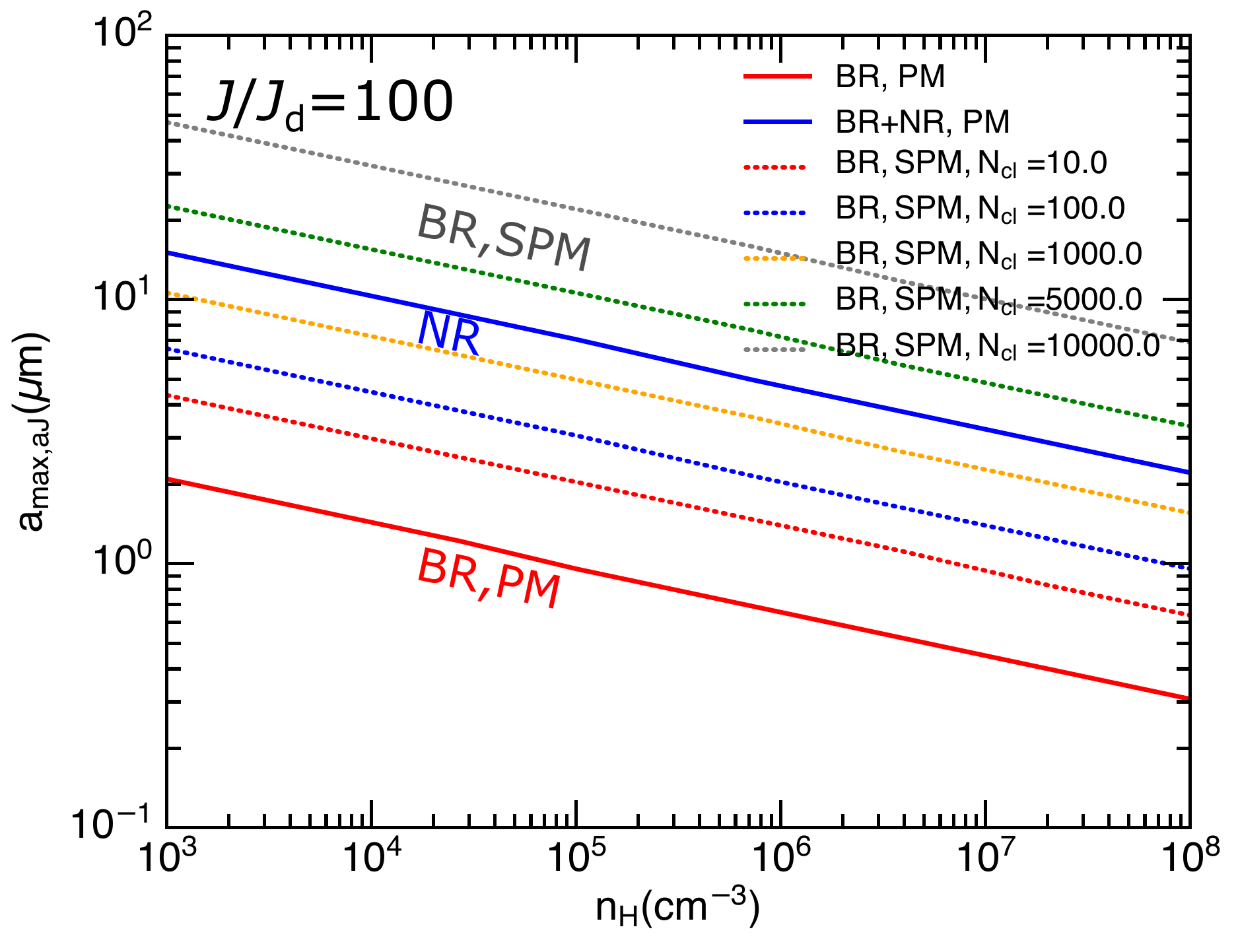}
\caption{Variation of the maximum grain size for efficient internal alignment ($a_{\max,aJ}$) due to Barnett relaxation and nuclear relaxation with the local gas density, assuming the thermal rotation (left panel) and suprathermal rotation (right panel). The values of $a_{\max,aJ}$ decrease with the gas density but increase with the level of iron inclusions ($N_{\rm cl}$). Suprathermal rotation also increases the internal alignment size as expected.} 
\label{fig:amax_align}
\end{figure*}

Figure \ref{fig:amax_align} (left panel) shows the decrease of $a_{\max,aJ}$ with the gas density numerically computed for paramagnetic and superparamagnetic grains with varying $N_{\rm cl}$, assuming $J/J_{d}=1$. The magnetic field strength as a function of the gas density is approximated as $B\approx 49(n_{\H}/10^{4}\cm^{-3})^{0.65}\mu G$ for $n_{\H}>300\cm^{-3}$, and $B\approx5\mu$G for $n_{\H}<300\cm^{-3}$ (\citealt{Crutcher:2010p318}). The results for ordinary paramagnetic material are also shown in solid lines. For the PM dust, Barnett relaxation induces the alignment of small grains only, with $a_{\max,aJ}<0.5\mu$ (red line), which is much smaller than that due to nuclear relaxation of $a_{\max,aJ}\lesssim 3\mum$ (blue line). The results for SPM grains are shown in dotted lines. Internal alignment occurs for grains up to a size of $a_{\max,aJ}\lesssim 10\mum$, assuming $N_{\rm cl}\lesssim 10^{4}$. For the typical density of a dense core of $n_{\H}=10^{5}\cm^{-3}$, one has $a_{\max,aJ}\sim 5\mum$ for $N_{cl}=10^{4}$. Figure \ref{fig:amax_align} (right panel) shows the results for suprathermal rotating grains with $J/J_{d}=100$. Suprathermal rotation enhances the values of $a_{\max,aJ}$ as implied by Equations (\ref{eq:aBar}) and (\ref{eq:a_nucl}), so that very large grains (VLGs) up to $10-50\mum$ can have efficient internal alignment for $N_{cl}=10^{4}$.

\subsection{External Alignment}
External alignment of the grain angular momentum ($\bJ$) with the magnetic field ($\Bv$) can be induced by Davis-Greenstein magnetic relaxation \citep{1951ApJ...114..206D}, radiative torques (\citealt{1976Ap&SS..43..291D}; \citealt{1996ApJ...470..551D}; \citealt{2007MNRAS.378..910L}; \citealt{Hoang:2008gb}), and mechanical torques (\citealt{2007ApJ...669L..77L}; \citealt{2018ApJ...852..129H}). In the following, we review the basic processes and formulae for reference (see \citealt{LAH15} for a review).

\subsubsection{Larmor precession}
The interaction of the grain magnetic moment (Eq. \ref{eq:muBar}) with the external static magnetic field, governed by the torque component $(d\bJ/dt).\hat{\phi}=\sin\xi d\phi/dt \hat{\phi}=[-\bmu_{\Bar}\times \bB].\hat{\phi}=-\mu_{\Bar}B\sin\xi\hat{\phi}$, causes the regular precession of the grain angular momentum around the magnetic field direction (see Figure \ref{fig:KBE_RAT}). The rate of such a Larmor precession denoted by $\tau_{\rm Lar}$, is given by
\bea
\tau_{\rm Lar}&=&\frac{2\pi}{|d\phi/dt|}=\frac{2\pi I_{\|}\Omega}{|\mu_{\Bar}|B}=\frac{2\pi |\gamma_{e}|I_{\|}}{\chi(0)VB},\nonumber\\
&=&0.84\frac{\hat{\rho} a_{-5}^{2}}{\hat{\chi}\hat{B}}~\yr,
\label{eq:tauB}
\ena
where $\hat{B}=B/5\mu$G and $\hat{\chi}=\chi(0)/10^{-4}$ are the normalized magnetic field and magnetic susceptibility, respectively. 

For superparamagnetic grains, the Larmor timescale, $\tau_{\rm Lar, sp}$, is given by the same formula but $\chi(0)$ is replaced by $\chi_{\rm sp}$ (Eq. \ref{eq:chi_sp}).

Above, for simplicity, we have assumed that internal relaxation is efficient so that the grain is spinning along the axis of maximum inertia with $\bJ=I_{\|}\bOmega$. This assumption usually holds for dust grains in MCs since internal relaxation is very faster than Larmor precession.

\subsubsection{Davis-Greenstein Relaxation: Paramagnetic and Superparamagnetic material}
Following \cite{1951ApJ...114..206D}, a rotating paramagnetic grain with the angular momentum $\bJ$ subject to an external magnetic field experiences magnetic dissipation of the grain rotational energy due to the existence of the rotating magnetization with respect to the grain body. This {\it Davis-Greenstein} magnetic relaxation eventually leads to the alignment of $\bJ$ with the magnetic field (see Figure \ref{fig:KBE_RAT}). The characteristic time of the magnetic relaxation is given by 
\bea
\tau_{\rm DG} &=& \frac{I_{\|}}{VK(\omega)B^{2}}=\frac{2\rho a^{2}}{5K(\omega)B^{2}},\nonumber\\
&\simeq & 2.4\times 10^{6}\hat{\rho}a_{-5}^{2}\hat{B}^{-2}\hat{K}^{-1} \yr,\label{eq:tau_DG}
\ena
where $\hat{K}=K(\omega)/(10^{-13}\s)$.

Plugging $K(\omega)$ (Eq. \ref{eq:kappa_para}) and $K_{\rm sp}(\omega)$ (Eq. \ref{eq:kappa_sp}) into Equation (\ref{eq:tau_DG}) one obtains the timescales of alignment by magnetic relaxation for paramagnetic and superparamagnetic material.

\subsubsection{Radiative Torque (RAT) Alignment}
\begin{figure}
\includegraphics[width=0.5\textwidth]{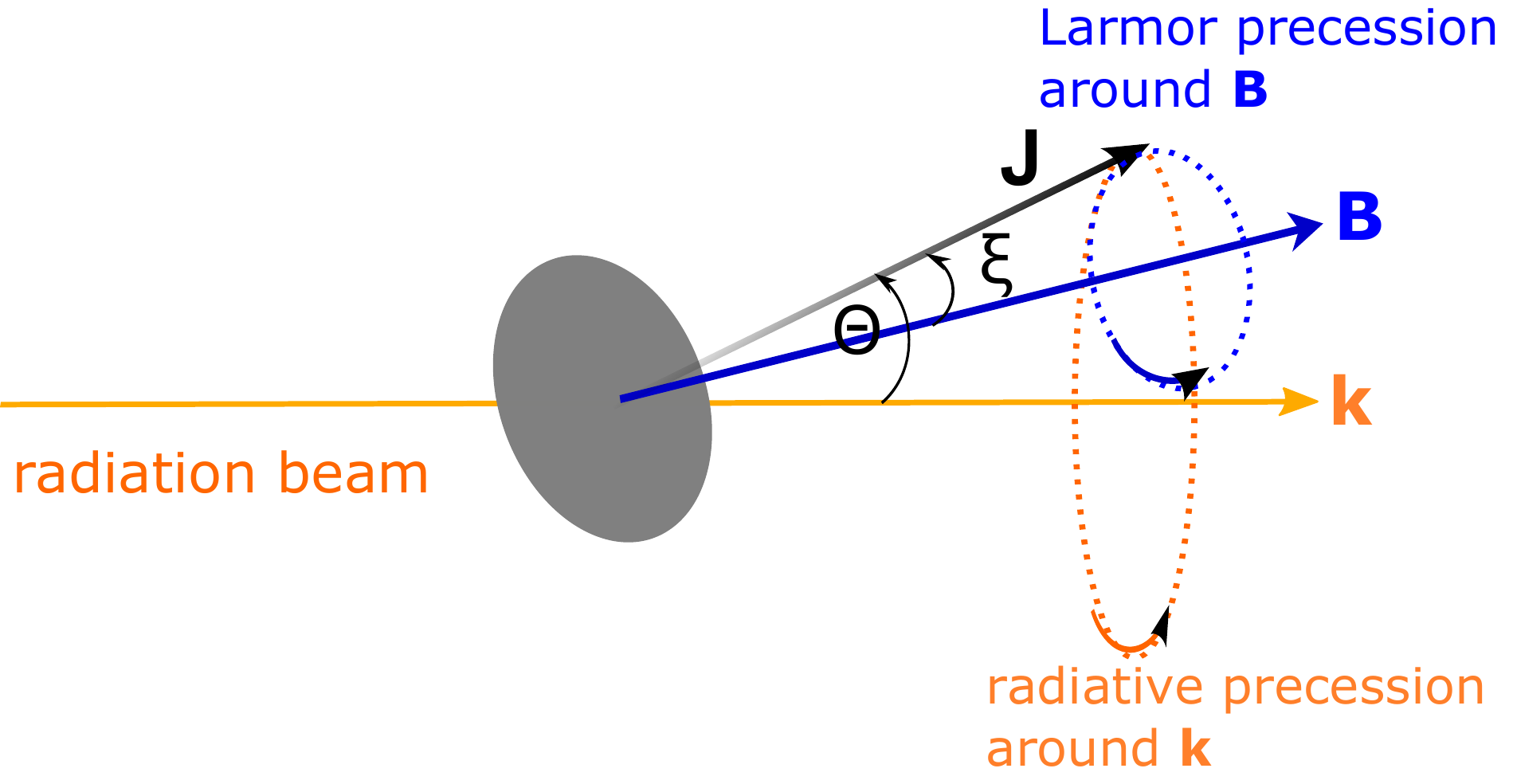}
\caption{Schematic illustration of the precession of the grain angular momentum (${\bf J}$) around the radiation direction (${\bf k}$) and the magnetic field (${\bf B}$), assuming the perfect internal alignment of the grain shortest axis with ${\bf J}$. The grain magnetic moment is directed along ${\bf J}$ due to Barnett effect. The fastest precession establishes the alignment axis by RATs.}
\label{fig:KBE_RAT}
\end{figure}

The interaction of an anisotropic radiation with an irregular grains induce radiative torques (RATs; \citealt{1976Ap&SS..43..291D}; \citealt{1996ApJ...470..551D}). RATs cause the grain precession around the radiation direction (${\bf k}$; see Figure \ref{fig:KBE_RAT}). Grains can be spun-up by RATs to angular velocities above its thermal value, so-called suprathermal rotation. Due to grain suprathermal rotation, efficient internal relaxation (Barnett and nuclear effects) quickly induces the alignment of the grain axis of maximum inertia with its angular momentum. Modern theory based on RATs (\citealt{2007MNRAS.378..910L}; \citealt{Hoang:2008gb}; \citealt{2016ApJ...831..159H}) implies that grain alignment with its angular momentum parallel to the magnetic field, resulting in the alignment of the shortest axis with the magnetic field when the Larmor precession is faster than the radiative precession. A detailed discussion of RAT alignment can be found in the review by \cite{LAH15}, here we provide some essential formulae for reference.

The radiative precession time around the radiation direction ($\bk$) by RATs is given by (\citealt{2007MNRAS.378..910L}; \citealt{2014MNRAS.438..680H}):
\bea
\tau_{k}&=&\frac{2\pi}{|d\phi/dt|}=\frac{2\pi J}{\gamma u_{\rad}\lambda a_{\rm eff} ^{2}Q_{e3}},\nonumber\\
&\simeq& 731
\hat{\rho}^{1/2}T_{1}^{1/2}\left(\frac{1.2\mum}{\gamma_{-1}\bar{\lambda}\hat{Q}_{e3}U}\right)\hat{s}^{1/6}a_{-5}^{1/2}\left(\frac{J}{J_{T}}\right)
\yr,\label{eq:tauk}
\ena 
where $\hat{Q}_{e3}=Q_{e3}/10^{-2}$ with $Q_{e3}$ the third component of RATs that induces the grain precession around $\bk$ (see Table 1 in \citealt{LAH15}). Comparing $\tau_{k}$ to the Larmor precession time, it can be seen that the Larmor precession is always faster than the radiation precession. Thus, the magnetic field is the axis of grain alignment (see Figure \ref{fig:KBE_RAT}).

The time required for RATs to spin-up grains to suprathermal rotation threshold, which is considered the RAT alignment time, is given by
\bea
\tau_{\rm RAT}&\equiv& \frac{3I_{\|}\Omega_{\rm T}}{\Gamma}= \frac{3\tau_{\rm gas}}{J_{\rm RAT}/J_{T}},\nonumber\\
&\simeq& \frac{2\times 10^{3}a_{-5}n_{3}^{-1}T_{1}^{1/2}yr}{(J_{\rm RAT}/3J_{\rm T})},\label{eq:tRAT}
\ena
$\Gamma$ is the radiative torque magnitude (\citealt{1996ApJ...470..551D}; \citealt{2007MNRAS.378..910L}), and $J_{\rm RAT}=\Gamma \tau_{\rm gas}$ is the maximum angular momentum spun up by RATs. For grains of $a=a_{\rm align}$ with $J_{\rm RAT}=3I_{\|}\Omega_{T}$, the RAT alignment timescale is comparable to $\tau_{\rm gas}$, and for $a>a_{\rm align}$, the RAT alignment is then much shorter because those grains rotate suprathermally with $J_{\rm RAT}>3J_{\rm T}$.

Numerical simulations show that grains can be efficiently aligned by RATs when they can rotate suprathermally (\citealt{Hoang:2008gb}; \citealt{2016ApJ...831..159H}). Taking the suprathermal condition that corresponds to the maximum grain angular momentum spun-up by RATs, $J_{\rm RAT}$, equal three times of its thermal value (i.e., $J_{\rm RAT}=3J_{T}$), one can determine the minimum size of grains that can be efficiently aligned by RATs. For a starless MC, \cite{Hoang.2021} derived the minimum effective size of aligned grains by interstellar radiation field (ISRF) as follows:
\bea
a_{\rm align}&=&\left(\frac{1.2n_{\rm H}T_{\rm gas}}{\gamma u_{\rm rad}\bar{\lambda}^{-2}} \right)^{2/7}\left(\frac{15m_{\rm H}k^{2}}{4\rho}\right)^{1/7}\nonumber\\
&\simeq &0.055\hat{\rho}^{-1/7} \left(\frac{\gamma_{-1}U}{n_{3}T_{1}}\right)^{-2/7}\left(\frac{\bar{\lambda}}{1.2\mum}\right)^{4/7}~\mum,\label{eq:aalign_ana}
\ena 
where $U=u_{\rm rad}/u_{\rm MMP83}$ is the ratio of the radiation energy density $u_{\rm rad}$ relative to that of the local ISRF in the solar neighborhood, $u_{\rm MMP83}$ (\citealt{1983A&A...128..212M}), $\bar{\lambda}$ is the mean wavelength of the local radiation field, $\gamma_{-1}=\gamma/0.1$ with $\gamma$ the anisotropy of the radiation field, and the damping due to infrared emission has been disregarded for starless dense clouds.

Equation (\ref{eq:aalign_ana}) implies $a_{\rm align} \sim 0.055, 0.21, 0.4 \mum$ for a dense cloud of $n_{\H}=10^{3}, 10^{5}, 10^{6}\cm^{-3}$ exposed to the local radiation field of $\gamma=0.1$, $U=1$, and $\bar{\lambda}=1.2\mum$. Accounting for the attenuation of ISRF toward the MC center, calculations in \cite{Hoang.2021} find that, at $A_{V}\sim 20$, the grain alignment size increases to $a_{\rm align}\sim 0.3, 0.7, 1.5\mum, $  for $n_{\H}\sim 10^{4}, 10^{5}, 10^{6}\cm^{-3}$. Therefore, the grain coagulation to micron sizes in dense MC essentially involves aligned grains. For grains smaller than $a_{\rm align}$, the coagulation can first occur for random grains and then become aligned by RATs when becoming larger than $a_{\rm align}$.

\subsubsection{Mechanical Torque (MET) Alignment}
Grains drifting through the gas experience mechanical torques (METs). As shown in \cite{2007ApJ...669L..77L} and \cite{2018ApJ...852..129H}, METs can efficiently align grains with the magnetic field, in analogy to RATs, i.e., the grain's shortest axis is parallel to the magnetic field when internal relaxation is efficient. For inefficient internal relaxation, grain alignment can occur with the long axis parallel to the magnetic field. However, the efficiency of METs is uncertain due to its complicated dependence on the grain shapes (\citealt{2018ApJ...852..129H}). Further study on METs is required to have an accurate assessment on METs (\citealt{Reissl.2022}). One expects that the RAT alignment can work in a wider range of environments than METs due to the ubiquity of radiation fields. However, METs may be important in dense clouds and the midplane of protoplanetary disks where radiation field is weak and grain drift is important .

\subsubsection{Maximum grain size for external alignment with the magnetic field}

\begin{figure*}
\centering
\includegraphics[width=0.49\textwidth]{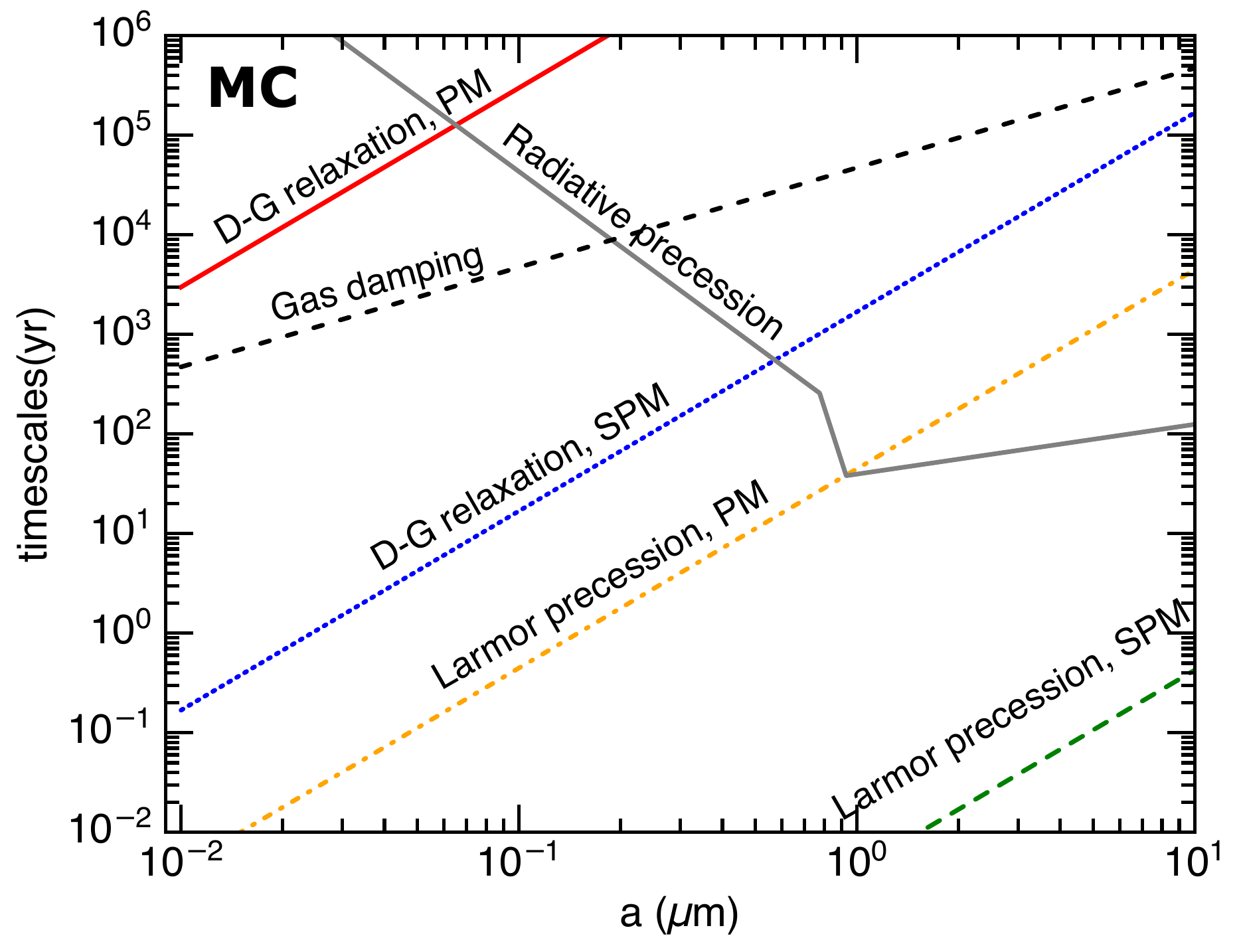}\includegraphics[width=0.49\textwidth]{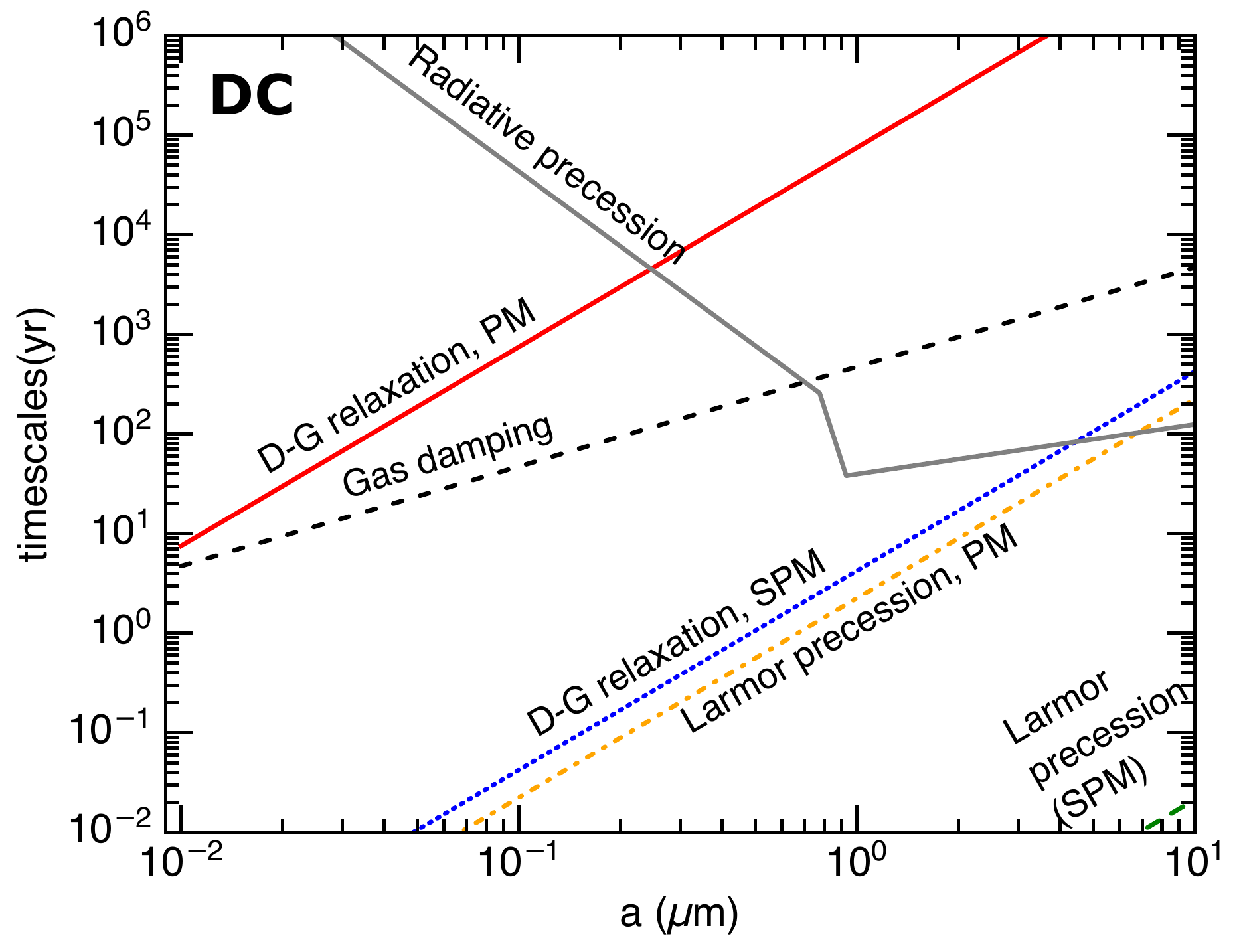}
\caption{Characteristic timescales of various processes in external grain alignment, including Davis-Greenstein relaxation, Larmor precession, radiative precession, and the gas damping time, for the MC of $n_{\H}=10^{3}\cm^{-3}$ (left panel) and DC of $n_{\H}=10^{5}\cm^{-3}$ (right panel). Both paramagnetic (PM) and superparamagnetic material (SPM) with iron inclusions of $N_{\rm cl}=10^{3}$ are considered. D-G paramagnetic relaxation is slower than the gas damping, but D-G superparamagnetic relaxation is much more efficient. Larmor precession is faster than radiative precession, and it has a shortest timescale for superparamagnetic grains for which the grain alignment occurs with respect to the magnetic field.}
\label{fig:talign_external}
\end{figure*}

Figure \ref{fig:talign_external} shows the timescales versus the grain size for various physical processes involved in external alignment, including Larmor precession and D-G relaxation for ordinary paramagnetic (PM) and superparamagnetic (SPM) material, radiation precession, and the gas damping for the MC (left panel) and DC (right panel) with physical parameters shown in Table \ref{tab:MC}. The radiation precession time by RATs (Eq. \ref{eq:tauk}) is calculated for the typical values of $\gamma=0.3$, $\bar{\lambda}=1.2\mum$, and $U=1$).

The Larmor precession of superparamagnetic grains is faster than the radiation precession, which implies the magnetic field is an axis of grain alignment instead of the radiation. The timescale for D-G relaxation for paramagnetic grains is longer than the gas damping time. However, superparamagnetic grains have the D-G relaxation time much shorter than the gas damping time such that the joint action of D-G relaxation and RATs can make the perfect alignment of $\bJ$ with $\Bv$ (\citealt{2016ApJ...831..159H}). 

We now estimate the maximum size of grains that still have external alignment with the magnetic field. Using Equations (\ref{eq:tgas}) and (\ref{eq:tauB}), one can write the ratio of the Larmor precession to gas damping time for superparamagnetic grains as:
\bea
\frac{\tau_{\rm Lar,sp}}{\tau_{\gas}}&=&\frac{2\pi g_{e}\mu_{B}}{\hbar}\left(\frac{1.2\sqrt{\pi}n_{\H}m_{\H}v_{\rm T}\Gamma_{\|}a}{\chi_{\sp}(0)sB}\right),\nonumber\\
&\simeq& 4.6\times 10^{-5}\left(\frac{n_{5}T_{d,1}T_{2}^{1/2}a_{-5}\Gamma_{\|}}{N_{\rm cl}\phi_{\rm sp,-2}B_{3}}\right),
\ena
where $n_{5}=n_{\H}/10^{5}\cm^{-3}$, $\phi_{\rm sp,-2}=\phi_{\rm sp}/0.01$, and $B_{3}=B/1000\mu$G. Above, we have assume $K(\omega)=\chi_{2}(\omega)/\omega\approx \chi(0)$ which is valid for large grains of $\omega\tau_{2}\ll 1$.

The maximum size for the grain alignment with $\bJ$ aligned with the magnetic field ($\Bv$), denoted by $a_{\max, JB}$, is then determined by $\tau_{\rm Lar,sp}/\tau_{\gas}=1$, yielding
\bea
a_{\max, JB} = 1.4\times 10^{3} \left(\frac{N_{\rm cl,4}\phi_{\rm sp,-2}B_{3}}{n_{5}T_{1}^{1/2}\Gamma_{\|}}\right) \cm,\label{eq:aLar}
\ena
which implies that VLGs of $1000$ cm can still be aligned with the magnetic field in dense clouds of $n_{\H}\sim 10^{5}\cm^{-3}$. However, in the region of high density $n_{\H}\sim 10^{12}\cm^{-3}$ like the midplane of protoplanetary disks (PPDs), only grains upto $a_{\max,JB}\sim 1\mum$ can be aligned with the magnetic field, assuming $N_{\rm cl}\sim 10^{4}$.

In summary, our main results in this section suggest that composite dust grains of size $a\sim [a_{\rm align}-a_{\max,aJ}]$ have efficient internal alignment and external alignment with the magnetic field, whereas VLGs of sizes $a\sim [a_{\max,aJ}-a_{\max,JB}]$ have inefficient internal alignment but efficient external alignment with the magnetic field. Even in protostellar cores of high density $n_{\H}\sim 10^{7}-10^{8}\cm^{-3}$, composite grains with iron inclusions can have both efficient internal alignment and external alignment up to $a_{\max,aJ}\sim 10\mum$ due to internal relaxation, leading to the alignment of the grain's shortest axis with the angular momentum and then with the magnetic field. However, VLGs of size $a\sim [\sim 10\mum-1\cm]$ only have efficient external alignment but inefficient internal alignment, which may induce the alignment with their shortest axes perpendicular to the angular momentum and magnetic field, (i.e., {\it wrong alignment}) (\citealt{2009ApJ...697.1316H}). 

\section{Grain growth by gas accretion to drifting aligned grains}\label{sec:accretion}
In this section, we discuss the effect of grain alignment with the magnetic field on grain growth from submicron to micron sizes in MCs by gas accretion and grain-grain collisions.

Suppose that a randomly-oriented, tiny grain of size $a<a_{\rm align}$ drifts through the interstellar gas with a velocity $v_{d}$. The gas includes H, He, and heavy (i.e., metal) elements (X) with $X=C, O, Mg, Si, Fe$, etc. In MCs, the accretion of H leads to the formation of $H_2$ and a water ice mantle, and part of $H_2$ molecules rapidly desorb from the grain surface. Accretion of metal elements and the ice mantle buildup gradually increase the grain size to $a>a_{\rm align}$. Subsequently, gas accretion acts on the grain that is aligned with the shortest axis parallel to the magnetic field.

\subsection{Non-aligned grains}
We first consider the accretion of the gas to a tiny, randomly oriented grain. Since the drift velocity of tiny grains by MHD turbulence is small (subsonic), one can ignore the drift. Due to randomization by gas collision, the gas accretion to the grain leads to isotropic increase in the grain size. The accretion rate of heavy (metal) atoms of element $i$ to the grain is then given by the same formulae as the gas accretion onto a spherical grain:
\bea
\frac{dm_{\rm gr}}{dt}=\sum_{X}m_{i} n_{i}\langle v\rangle_{i} \pi a^{2}S_{i}.\label{eq:accret}
\ena
where $S_{i}$ is the sticking coefficient, $\langle v\rangle_{i} =(8kT_{\rm gas}/\pi m_{i})^{1/2}\simeq 0.46T_{1}^{1/2}A_{i}^{-1/2}\km\s^{-1}$ where $T_{1}=T_{\rm gas}/10\K$ is the mean thermal speed and $m_{i}\approx A_{i}m_{H}$ with $A_{i}$ the atomic mass number, and the summation is carried over all heavy elements (X).

The increase in the grain size by gas accretion is
\bea
\frac{da}{dt}=\frac{\sum_{X}m_{i} n_{i}\langle v\rangle_{i} \pi a^{2}S_{i}}{4\pi a^{2}\rho}=\frac{m_{X} n_{X}\langle v\rangle_{X} \pi a^{2}S_{X}}{4\pi a^{2}\rho}.
\ena

The accretion time to increase the grain size by a thickness of $a$ is
\bea
\tau_{\rm acc}&=&\frac{a}{da/dt}=\frac{4a\rho}{\rho_{X}\langle v\rangle_{X}S_{X}}=\frac{4a\rho}{X\rho_{\rm gas}\langle v\rangle_{X} S_{X}}\nonumber\\
&\simeq &1.6\times 10^{7}\left(\frac{A_{X}}{20}\right)^{1/2}\left(\frac{\hat{\rho}a_{-5}}{n_{3}T_{1}^{1/2}X_{-2}S_{X}}\right)~\yr,\label{eq:tacc}
\ena
where the ratio of metal to gas mass density, $X= \rho_{X}/\rho_{\rm gas}$, with $\rho_{X}=m_{X}n_{X}$, $\rho_{\rm gas}=\mu m_{\rm H}n_{\rm H}$ with $\mu=1.4$ the mean molecular weight of the gas with $10\%$ mass from He, and $A_{X}$ is the mean atomic mass of metals which is taken to be $A_{X}=20$ in our numerical calculations.

Equation (\ref{eq:tacc}) reveals that the grain can grow to twice its original size in $\sim 30$ Myr, assuming the sticking coefficient $S_{X}=0.5$ and $n_{\H}=10^{3}\cm^{-3}$. However, the grain growth by gas accretion is constrained by the metal budget in the gas and destruction processes such as shattering and rotational disruption by RATs (\citealt{Hoang:2019da}).

\subsection{Aligned grains}
We now discuss the gas accretion to aligned grains which have sizes $a_{\rm align}<a<a_{\max,aJ}$. We assume that the aligned grain drifts through the gas with the velocity $v_{d}$ along the $\xhat$ direction, so ${\bf v}_{d}=v_{d}\hat{x}$. Let $\hat{x}\hat{y}\hat{z}$ be the lab frame in which $\hat{z}$ is chosen to be along the magnetic field, ${\bf B}$.

Due to fast spinning along the shortest axis, the grain can be approximated as an oblate spheroid with the lengths of the semi-major and minor axes denoted by $a$ and $c<a$, respectively (see Section \ref{sec:physics}). For simplicity, we assume that grain alignment is perfect with $\ahat_{1}\|\bJ\|\Bv$. Because the rotation period is much shorter than the mean time between two gas-grain collisions, the geometrical cross-section of the frontal surface is $\pi ac$ and of the perpendicular surface $\pi a^{2}$. 

The collision rate of the gas metal $X$ to the grain along the direction of the grain motion is (see Appendix \ref{sec:apdx}):
\bea
F_{x}&=&\frac{1}{4}n_{X} \langle v\rangle_{X} \left(e^{-s_{d}^{2}} +\sqrt{\pi}s_{d}[1+\erf(s_{d})]\right),~\label{eq:Rx}
\ena
where
\bea
s_{d}=\frac{v_{d}}{v_{T,X}}=1.1T_{1}^{-1/2}\left(\frac{v_{d}}{0.1\km\s^{-1}}\right)\left(\frac{A_{X}}{20}\right)^{1/2}\label{eq:sdX}
\ena
with $v_{T,X}=(2kT_{\rm gas}/A_{X}m_{\H})^{1/2}$ the thermal velocity of the metal $X$ in the gas. The first term in Equation (\ref{eq:Rx}) describes the reduction effect from trailing collisions due to grain motion, while the second term describes the head-on collisions. 

The collision rates along the perpendicular direction to the grain motion are
\bea
F_{y}&=&\frac{1}{4}n_{X}\langle v\rangle_{X},\label{eq:Ry}\\
F_{z}&=&\frac{1}{4}n_{X}\langle v\rangle_{X}.\label{eq:Rz}
\ena

The ratio of the collision rates along the two parallel and perpendicular directions to the grain motion is  
\bea
\frac{F_{x}}{F_{y}}=\frac{F_{x}}{F_{z}}=\left(e^{-s_{d}^{2}} +\sqrt{\pi}s_{d}[1+\erf(s_{d})]\right).~~\label{eq:Fx_Fy}
\ena

\begin{figure}
\includegraphics[width=0.5\textwidth]{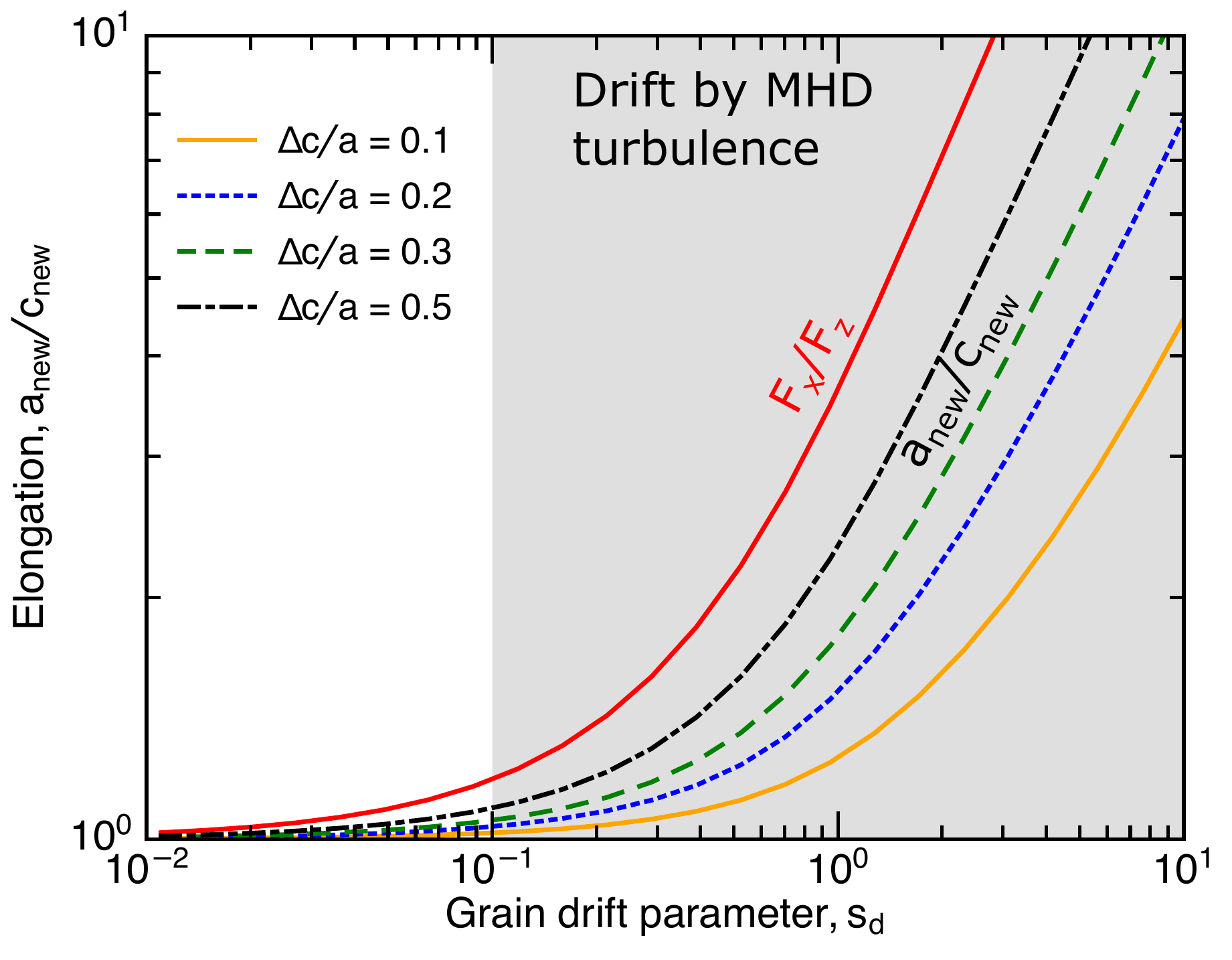}
\caption{The ratio of the collision rate along the grain motion to its perpendicular direction ($F_{x}/F_{z}$) and the grain elongation as functions of the dimensionless parameter of the grain drift relative to the metals, $s_{d}=v_{d}/v_{T,X}$ (Eq. \ref{eq:sdX}) when the grain radius is increased by $\Delta c/a=0.1-0.5$. Transonic and supersonic grains become highly elongated with elongation greater than $\sim 3$ by the time the grain size accumulates $50\%$ of its radius. The shaded area marks the possible range of grain drift enabled by MHD turbulence in MCs.}
\label{fig:RxRy}
\end{figure}

For $s_{d}=0$, one has $F_{x}/F_{y}=1$, which is the case of isotropic gas accretion. For $s_{d}\gg 1$, the second term of Equation (\ref{eq:Fx_Fy}) dominates such that $F_{x}/F_{y}\rightarrow 2\sqrt{\pi}s_{d}$, which implies that gas accretion is mostly along the direction of the grain motion. At the drift parameter of $s_{d}=1$, the head-on collision rate is about four times greater than the collision rate along the magnetic field (see Figure \ref{fig:RxRy}, red line). 

The rate of the increase in the grain mass by accretion of metals along the direction of the grain motion is
\bea
\frac{dm_{x}}{dt}&=&A_{X} m_{\rm H} F_{x}\pi acS_{X}\nonumber\\
&=&\frac{1}{4}S_{X}n_{X}A_{X} m_{\rm H}\pi ac \langle v\rangle_{X} \left(e^{-s_{d}^{2}} +\sqrt{\pi}s_{d}[1+\erf(s_{d})]\right).~~~~~
\ena
where $S_{X}$ is the sticking coefficient of metals.

The rate of the increase in the grain mass by accretion along the direction perpendicular to the direction of the grain motion is
\bea
\frac{dm_{z}}{dt}=A_{X} m_{\rm H} F_{z}\pi a^{2}S_{X}=\frac{1}{4}A_{X} m_{\rm H}S_{X}n_{X}\pi a^{2}\langle v\rangle_{X}.
\ena

Assuming that the initial grain shape is slightly elongated with $a\approx c$, the new axial ratio of the grain after a time $\Delta t$ is then
\bea
\frac{a_{\rm new}}{c_{\rm new}}=\frac{a+\Delta a}{c+\Delta c}\approx \frac{1+\Delta a /a}{1+\Delta c/a}.
\ena
For $\Delta c/a\ll 1$, then, $1/(1+\Delta c/a)\approx 1-\Delta c/a$. Thus,
\bea
\frac{a_{\rm new}}{c_{\rm new}}&=&\left(1+\frac{\Delta a}{a}\right)\left(1-\frac{\Delta c}{a}\right)\nonumber\\
&\approx&1+\frac{\Delta c}{a}\left(\frac{\Delta a}{\Delta c}-1\right)=1+\frac{\Delta c}{a}\left(\frac{F_{x}}{F_{z}}-1\right).~~
\ena
It can be seen that for $v_{d}=0$, the axial ratio is kept constant. For $v_{d}>0$, the axial ratio increase because $F_{x}/F_{z}>1$. By the time that the gas accretion increases the grain size by $\Delta c=a$, the axial ratio is increased to $a_{\rm new}/c_{\rm new}=(F_{x}/F_{z})$. 

Figure \ref{fig:RxRy} shows the increase of the grain elongation due to accretion of metals with an average mass number $A_{X}=20$ with the drift parameter at four epochs when the grain accumulates a surface layer of $10-50\%$ of its original radius. Transonic and supersonic grains become highly elongated with axial ratio of $a_{\rm new}/c_{\rm new}>2$ by the time the grain size accumulates $50\%$ of its radius. For the silicate grains accelerated by MHD turbulence in MCs with $v_{d}\sim 0.5\km\s^{-1}$ (see Table \ref{tab:MC}), which corresponds to $s_{d}\sim 5.5$ (see Eq.\ref{eq:sdX}), Figure \ref{fig:RxRy} implies an elongation of $a/c\approx 3 $ and $5$ by the time its radius increases by $10\%$ and $20\%$, respectively. Therefore, the grain shape becomes highly elongated after the grain radius is increased by just $10\%$. 
It is worth noting that the realistic elongation of an individual grain grown from gas accretion is affected by the metal abundance in the gas, grain shattering, and rotational disruption.



\begin{table*}[htb]
\begin{center}
\caption{\label{tab:timescales}
Timescales of physical processes involved in grain alignment and grain growth}
\begin{tabular}{l l }
\hline\hline
Definitions and Physical processes & Typical Values\cr
\hline
Oblate spheroidal grain shape and axial ratio  & $a\times a\times c$, $s=c/a$\cr
Gas damping time, $\tau_{\rm gas}(\yr)$ &$8.3\times 10^{3}sa_{-5}\hat{\rho}\left(\frac{1}{n_{3}T_{1}^{1/2}}\right)$\cr
Barnet relaxation (paramagnetic), $\tau_{\rm BR}(\yr)$ & $0.5\hat{\rho}^{2}f(\hat{s})a_{-5}^{7}\left(\frac{J_{d}}{J}\right)^{2}
\left[1+\left(\frac{\omega\tau_{2}}{2}\right)^{2}\right]^{2} $  \cr 
Nuclear relaxation, $\tau_{\rm NR}(\s)$& $125\hat{\rho}^2 f(\hat{s})a_{-5}^{7}\left(\frac{J_{d}}{J}\right)^2\left(\frac{n_e}{n_n}\right) \left(\frac{g_n}{3.1}\right) \left(\frac{2.7 \mu_N}{\mu_n}\right) \left[1+\left(\frac{\omega \tau_n}{2} \right)^2 \right]^{2}$\cr
Barnett relaxation (superparamagnetic),$\tau_{\rm BR,sp}(\yr)$ & $3.2 \hat{\rho}^{2}f(\hat{s})a_{-5}^{7}\frac{1}{N_{\rm cl}}\left(\frac{J_{d}}{J}\right)^{2}\times \left[1+\left(\frac{\omega\tau_{\rm sp}}{2}\right]^{2}\right)^{2} $  \cr 
D-G relaxation, $\tau_{\rm DG}(\yr)$ &$2.4\times 10^{6}\hat{\rho}a_{-5}^{2}\left(\frac{B}{5\mu G}\right)^{-2}\left(\frac{10^{-13}\s}{K(\omega)}\right)  $  \cr
Larmor precession, $\tau_{\rm Lar}(\yr)$ & $8.4\hat{\rho}\hat{\chi}^{-1}\hat{B}^{-1} a_{-5}^{2}~\yr$ \cr
Radiative precession, $\tau_{k}(\yr)$ & $731
\hat{\rho}^{1/2}T_{1}^{1/2}\left(\frac{1.2\mum}{\gamma_{-1}\bar{\lambda}\hat{Q}_{e3}U}\right)\hat{s}^{1/6}a_{-5}^{1/2}\left(\frac{J}{J_{T}}\right)$ \cr
Gas accretion time, $\tau_{\rm acc}(\yr)$ &$1.2\times 10^{6}\frac{\hat{\rho}a_{-5}}{n_{3}T_{1}^{1/2}X_{-2}}$  \cr
Grain-grain collision, $\tau_{\rm gg}(\yr)$ &$7.6\times 10^{4}\hat{\rho}a_{-5}n_{3}^{-1}\left(\frac{v_{d}}{1\rm km\s^{-1}}\right)^{-1}$  \cr
External alignment (RAT) time, $\tau_{\rm RAT}(\yr)$ &$2\times 10^{3}\frac{a_{-5}n_{3}^{-1}T_{1}^{1/2}}{(J_{\rm RAT}/3J_{T})} $  \cr
Maximum size of magnetic alignment ($\bJ$ with $\Bv$),$a_{\max, JB}(\cm)$ & $ 1.4\times 10^{3} \left(\frac{N_{\rm cl,4}\phi_{\rm sp,-2}B_{3}}{n_{5}T_{1}^{1/2}\Gamma_{\|}}\right)$ \cr
Maximum size of internal alignment ($\ahat_{1}$ with $\bJ$), $a_{\max,aJ}(\mum)$ & $1.0h^{1/3} \left(\frac{N_{\rm cl,4}\phi_{\rm sp,-2}}{n_{5}T_{1}^{1/2}\Gamma_{\|}}\right)^{1/6}\left[\exp\left(\frac{N_{\rm cl}T_{\rm act}}{T_{d}}\right)\right]^{1/6}\times\left(\frac{1}{1+(\omega\tau_{\rm sp}/2)^{2}}\right)^{1/3}\left(\frac{J}{J_{d}}\right)^{1/3}$\cr
\hline
\multicolumn{2}{l}{Notes: $\hat{s}=s/0.5, a_{-5}=a/(10^{-5}\cm)$, $f(\hat{s})=\hat{s}[(1+\hat{s}^{2})/2]^{2}, n_{3}=n_{\H}/(10^{-3}\cm^{-3}),~T_{1}=T_{\rm gas}/10\K, \hat{\chi}=\chi(0)/10^{-4}$}\cr
\cr
\hline\hline\\
\end{tabular}
\end{center}
\end{table*}

\section{Grain coagulation from drifting aligned grains}\label{sec:growth}
In this section, we first estimate the timescale required for grain growth by grain collisions and compare with the characteristic timescales of grain alignment to see the relevance of the latter process. We then discuss the expected shape and structure of dust aggregates formed from the coagulation from aligned grains.

\subsection{Grain coagulation by collisions}

The mean time between two successive collisions of two equal-size grains is given by
\begin{eqnarray}
\tau_{\rm gg}&=&\frac{1}{\pi a^{2} n_{\rm gr}v_{d}}=\frac{4\rho a M_{g/d}}{3n_{\rm H}m_{\rm H}v_{d}}\nonumber\\
&\simeq& 7.6\times 10^{4}\hat{\rho}a_{-5}n_{3}^{-1}\left(\frac{v_{d}}{1\km\s^{-1}}\right)^{-1}{~\rm yr},
\end{eqnarray}
where $n_{\rm gr}$ is the number density of dust grains, $M_{g/d}= \mu n_{\H}m_{\H}/n_{\rm gr}m_{\rm gr}=100$ with $m_{\rm gr}=4\pi a^{3}\rho/3$ is the gas-to-dust mass ratio, and we have assumed the single-grain size distribution.


The shattering threshold is $v_{\rm shat}\sim 2.7\km\s^{-1}$ for silicate and $1.2\km\s^{-1}$ for graphite grains (\citealt{1996ApJ...469..740J}; \citealt{Yan:2004ko}). At high velocities, shock waves are produced inside the grains and can shatter them in smaller fragments. Grain velocities achieved by MHD turbulence in MC and DC are $v_{d}<v_{\rm shat}$. Therefore, grain shattering does not occur in the MC and DC. However, whether grains of $v<v_{\rm shat}$ stick upon collision is unclear, and the presence of ice mantles is expected to enhance sticking collisions due to its larger threshold $v_{cri}$. Therefore, in this paper, we assume that sticky collisions and grain coagulation occur whenever $v<v_{\rm shat}$.

\begin{figure}
\includegraphics[width=0.5\textwidth]{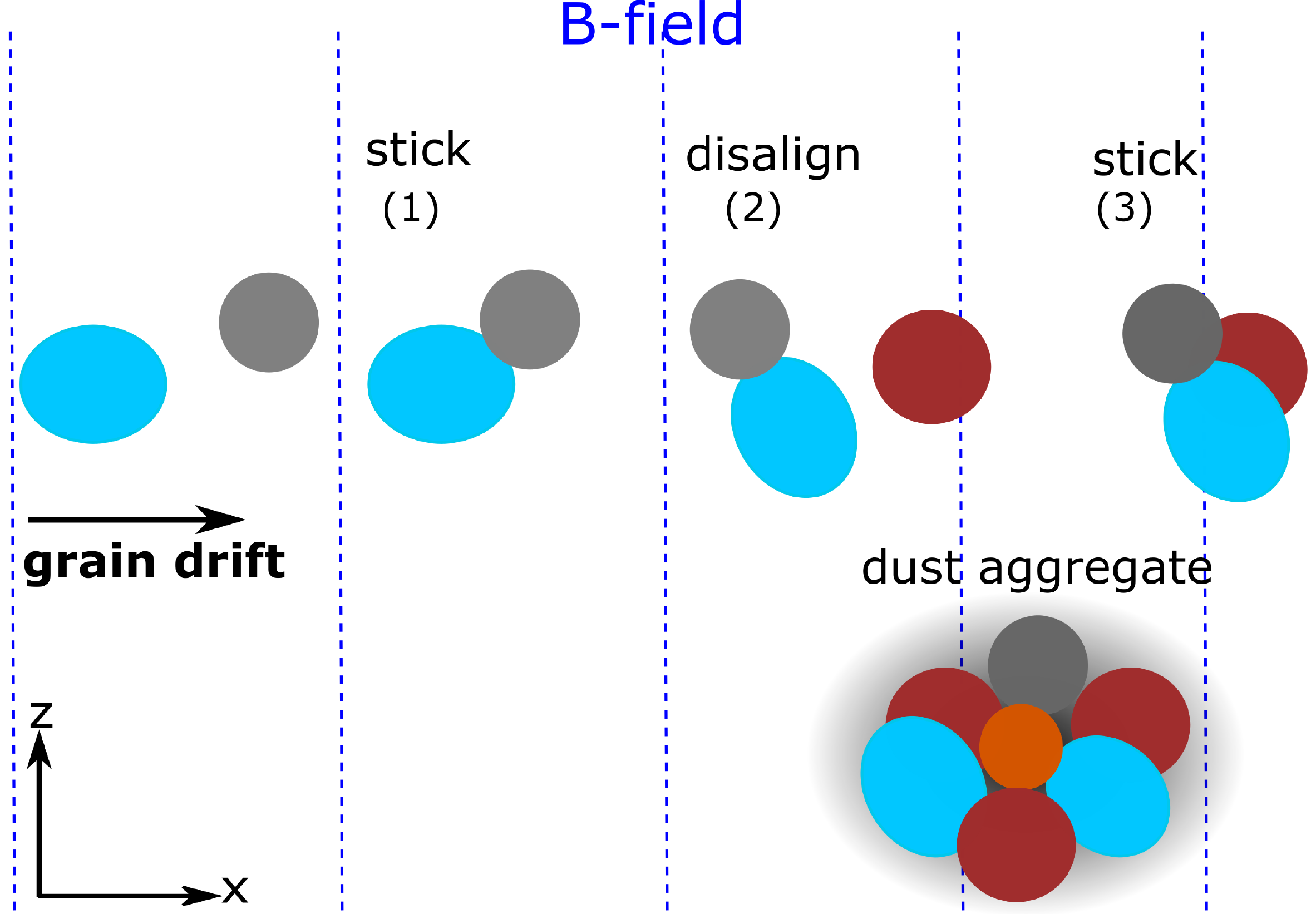}
\caption{Illustration of coagulation by collisions of randomly oriented grains moving along the x-axis, perpendicular to the mean magnetic field (vertical dashed lines). After the first collision, grains stick and form an elongated grain (stage 1). The orientation of the resulting binary is randomized by gas collisions (stage 2), and then experiences the next collision along the direction with maximum cross-section, resulting in the large grain (stage 3) and eventually a dust aggregate.}
\label{fig:growth_noalign}
\end{figure}

\begin{figure}
\includegraphics[width=0.5\textwidth]{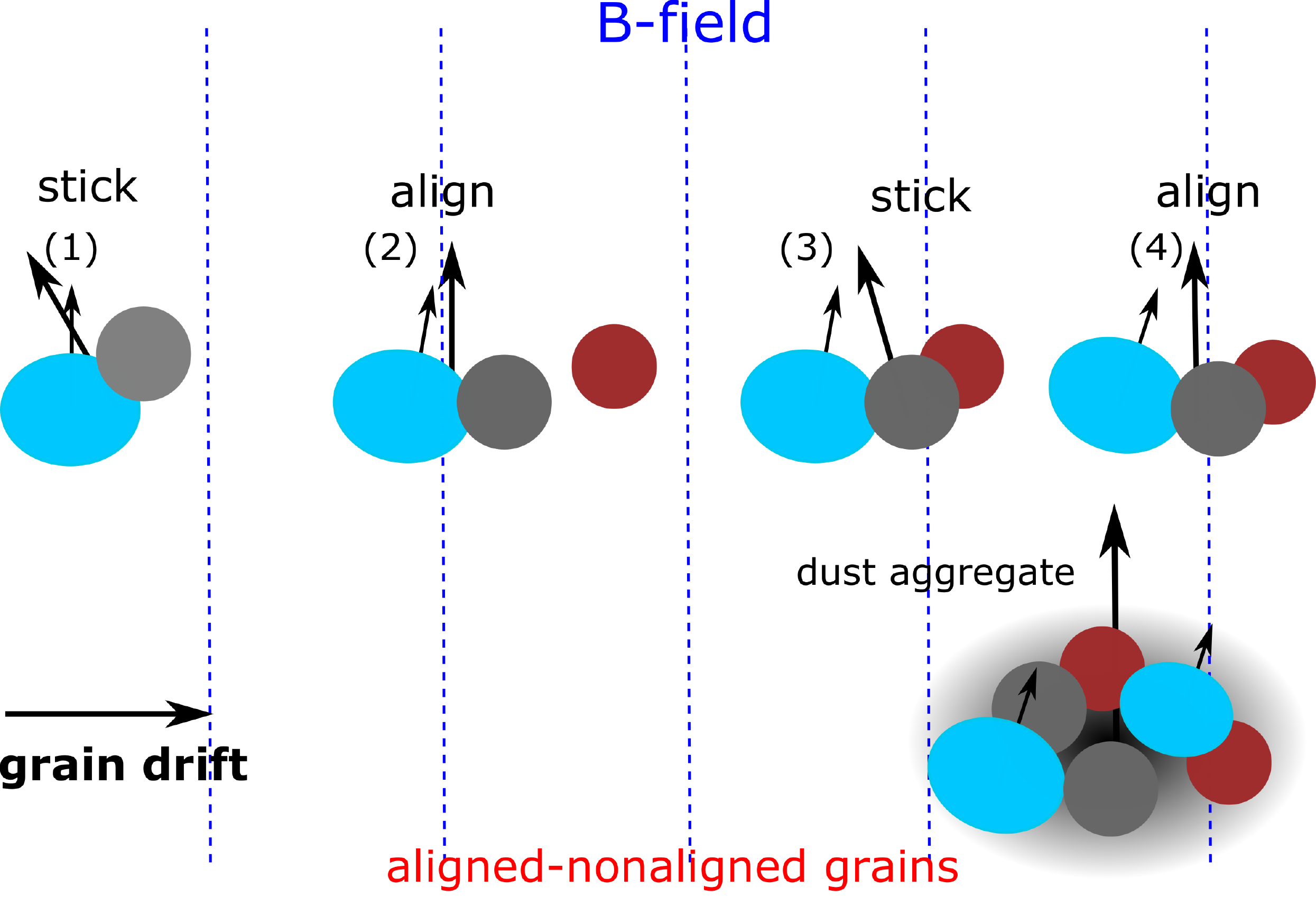}
\caption{Illustration of grain coagulation by collisions of aligned-unaligned grains moving perpendicular to the mean magnetic field (vertical dashed lines). On the grains, the short arrow denotes the shortest axis of the aligned grain, and the long arrow denotes the shortest axis of the composite grain. After the first collision, the grains stick to become a binary (stage 1). The resulting elongated grain is rapidly aligned with the long axis perpendicular to the magnetic field (stage 2). The subsequent collision occurs along the long axis, results in the more elongated grain shape (stage 3), and becomes aligned with the magnetic field (stage 4). Finally, an elongated dust aggregate is formed.}
\label{fig:growth_align_unalign}
\end{figure}

Table \ref{tab:timescales} summarizes the characteristic timescales involved in grain alignment and grain growth. The accretion and grain-grain collision time are much longer than the internal alignment time (Barnett relaxation) and external alignment processes (Larmor precession and RAT alignment). As shown, the alignment time by RATs is at most $\tau_{\rm RAT}\sim \tau_{\rm gas}$, which is equal to the time required to collide with the gas of the same mass as the dust grain. Since the dust mass is $\sim 1$ percent of the gas mass, the grain collision time is 100 times longer. Note that the collision time between the grain size $a$ with a tiny grain is shorter due to its higher density. However, such collisions do not randomize significantly the grain orientation because of its lower mass. Therefore, we can assume that aligned grains can be rapidly re-aligned before hitting another grain.

\subsection{Nonaligned-nonaligned grains: stick and disalign}
Consider first the collisions between randomly oriented grains. In this scenario, grain collisions are similar to what is studied previously. Figure \ref{fig:growth_noalign} illustrates the grain coagulation by collisions of non-aligned grains moving perpendicular to the mean magnetic field. After the first collision, grains stick and form an elongated binary. The newly formed grain is rapidly randomized by gas collisions, and then experiences the next random collision with the high probability along the direction with maximum cross-section. Therefore, coagulation of non-aligned grains leads to a dust aggregate of low elongation.

\subsection{Aligned-nonaligned grains: stick and align}
We now consider coagulation by collisions between an aligned grain with another nonaligned grain. Figure \ref{fig:growth_align_unalign} illustrates the coagulation of one aligned grain with non-aligned grains. After sticking collisions, the monomers stick to become a binary (stage 1). If the collision results in the sudden disalignment so that the shortest axis deviates significantly from the angular momentum which is parallel to the magnetic field, then, internal relaxation rapidly brings the grain axis to be aligned with the magnetic field, provided that the particle size is smaller than $a_{\max,aJ}$ (see Figure \ref{fig:amax_align}). Note that the alignment timescale is shorter than the grain collision time $\tau_{\rm gg}$ (see Table \ref{tab:timescales}). Thus, the resulting elongated binary can be rapidly realigned with the long axis perpendicular to the magnetic field (stage 2). The subsequent collision occurs along the long axis, resulting in a more elongated particle (stage 3), and the re-alignment occurs due to internal relaxation and RAT alignment (stage 4). A dust aggregate is finally formed, with a larger elongation than in the case of growth from non-aligned grains (see Figure \ref{fig:growth_noalign}). Moreover, in the RAT paradigm, the nonaligned grain is smaller than aligned grains, so the collision would not result in a significant deviation of the net grain angular momentum from the magnetic field. 

\subsection{Aligned-aligned grains: stick and align}
\begin{figure}
\includegraphics[width=0.5\textwidth]{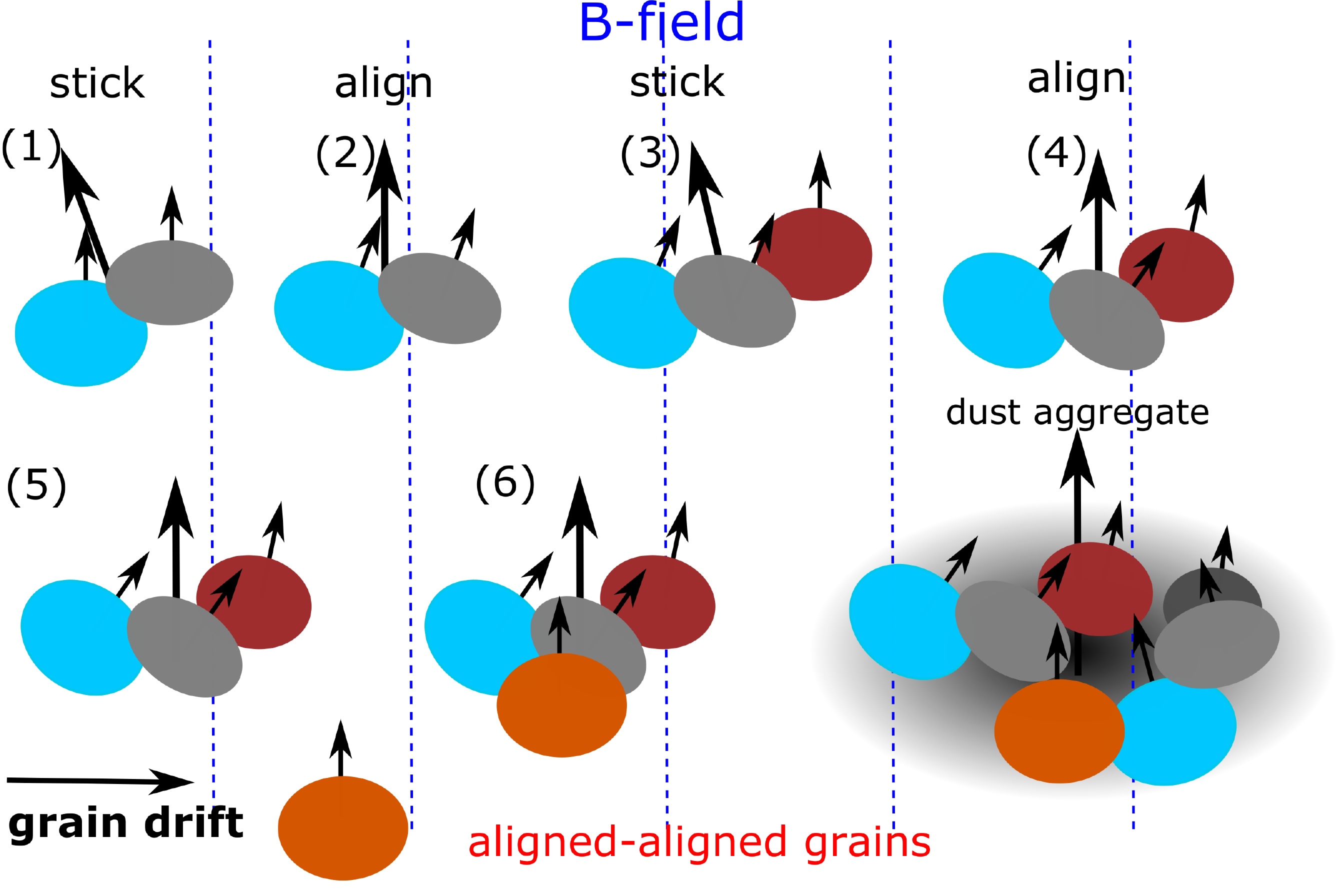}
\caption{Same as Figure \ref{fig:growth_align_unalign} but for collisions of aligned-aligned grains. The composite grain consists of elongated monomers with the shortest axes (short arrows) parallel, and the shortest axis of the composite grain (long arrow) is directed along the magnetic field. Finally, a dust aggregate is formed that contains elongated grains with parallel short axes.}
\label{fig:growth_align_align}
\end{figure}

Figure \ref{fig:growth_align_align} illustrates the coagulation from collisions between two aligned grains. After the first collision, a binary is formed and comprises a pair of aligned grains with parallel minor axes (short arrows). The minor axis of the binary (long arrow) is not necessary parallel to the grain's axes and makes a small angle (stage 1). However, internal relaxation rapidly brings the binary axis to be aligned with the magnetic field (stage 2). The first binary continues to collide with another aligned grain and coagulates, but the minor axes of the binary are slightly tilted from the third grain's axis (stage 3). The resulting particle has the shortest axis titled from the magnetic field and the angular momentum. The internal relaxation again acts to bring the internal alignment on a timescale of $\tau_{\rm Bar}$ (stage 4). An aligned grain moving along the magnetic field can collide with the large grain and forms a dust particle (stage 5), subsequently relaxation processes bring it to be aligned with the magnetic field (stage 6). A dust aggregate is finally formed. The shortest axis of the dust aggregate is tilted by some small angle with the short axes of the first binary. Thus, the coagulation from aligned grains leads to the formation of composite dust aggregates that contains elongated binaries made of oblate spheroids with parallel short axes inheriting from grain alignment with the magnetic field. The shortest axis of the dust aggregate is aligned along the magnetic field.

\begin{figure}
\includegraphics[width=0.5\textwidth]{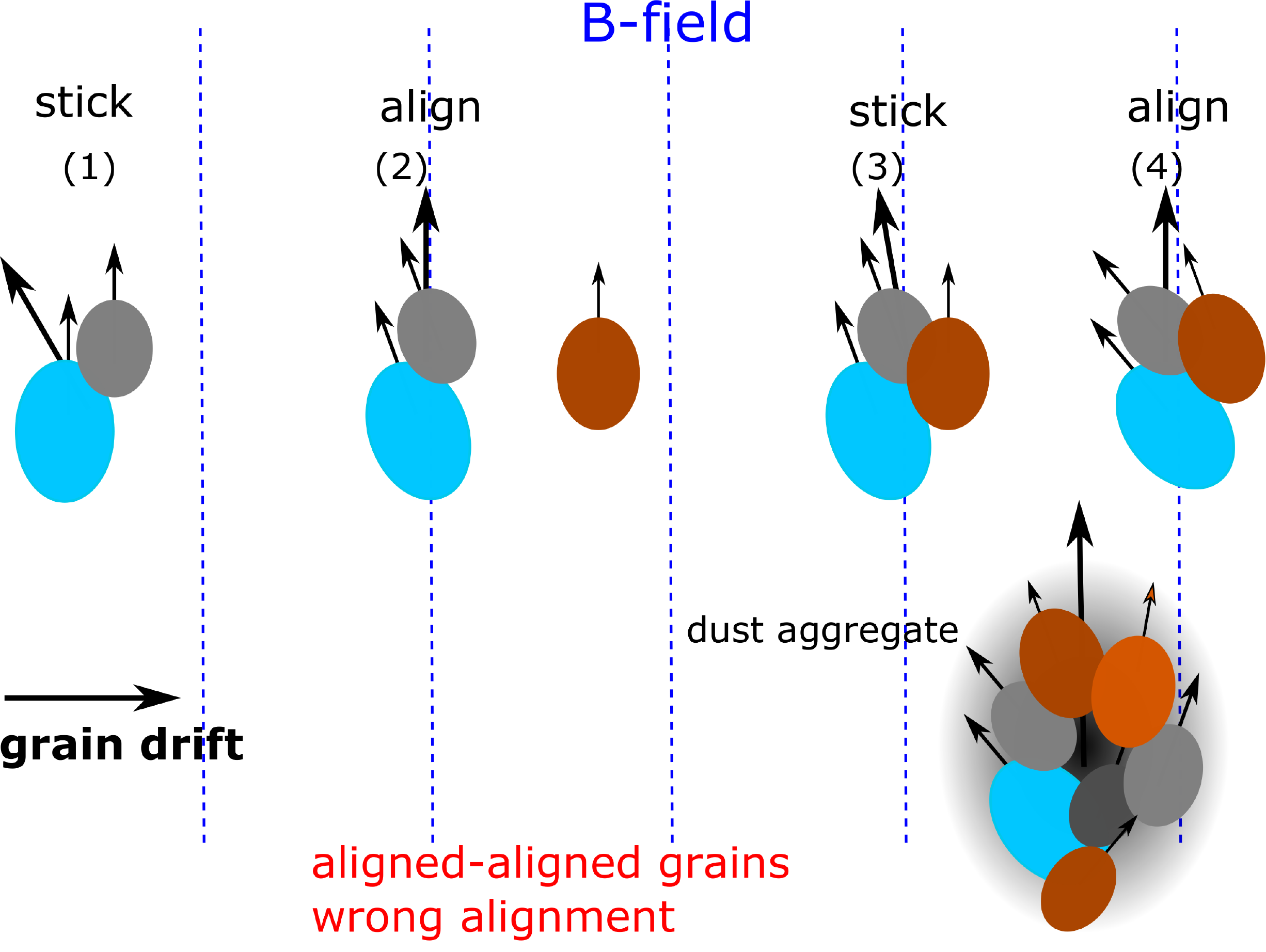}
\caption{Same as Figure \ref{fig:growth_align_align} but for grains with inefficient internal relaxation that induces the wrong alignment with the grain longest axis parallel to the spinning axis. Finally, a dust aggregate is formed that contains elongated grains with parallel long axes, and the long axis of the aggregate is parallel to the magnetic field due to wrong alignment.}
\label{fig:wrong}
\end{figure}

Figure \ref{fig:wrong} presents the coagulation of grains with wrong alignment with the long axis parallel to the magnetic field. The process is similar to coagulation of aligned grains in Figure \ref{fig:growth_align_align}, but the first binary is made of two aligned grains with parallel long axes. The final dust aggregate contains aligned grains with parallel long axes and has its long axis parallel to the magnetic field.

\section{Discussion}\label{sec:discuss}
\subsection{Grain shape and structures from grain growth of aligned grains}
We have studied the effect of grain alignment and motion across the magnetic field on grain growth by accretion of metals and grain-grain collisions. For grains aligned with the axis of maximum inertia (e.g., shortest axis) parallel to the magnetic field, the flux of gas species arriving along the direction of grain motion is increased significantly compared to the particle flux in the perpendicular direction (see Eq.\ref{eq:Fx_Fy}). As a result of gas accretion, the shape of grains becomes oblate spheroidal due to fast rotation of the grain around the magnetic field. 
In addition, the grain elongation by gas accretion increases over time and become extremely elongated of elongation $>5$ if the grain moves supersonically through the metals (see Figure \ref{fig:RxRy}). Interestingly, even with a low drift velocity of $v_{d}\sim 0.2\km\s^{-1}$, the dust grain drifts with $s_{d}\sim 2$, i.e.,  supersonically relative to metals due to the high atomic mass of metals (see Eq. \ref{eq:sdX}). However, the ultimate elongation of interstellar grains by gas accretion is also constrained by the metalicity available in the cloud and destruction processes such as grain shattering and rotational disruption by RATs (\citealt{Hoang:2019da}).

Coagulation from sticking collisions between aligned grains first produces a binary comprising two aligned grains of oblate shapes with parallel short axes. The binary grains can be rapidly aligned by efficient internal relaxation and RATs, and then continues to collide with another aligned grain, forming a bigger particle with aligned axes. The grain growth process continues and eventually forms a micron-sized composite dust aggregate (\citealt{1997ApJ...480..647D}; \citealt{Ormel:2009p4022}). Therefore, the dust aggregate contains elongated binaries of two aligned particles with different sizes and orientations, forming a hierarchy structure. 

Due to the rapid decrease in the efficiency of internal relaxation by Barnett effect with the grain size, as $a^{7}$ (see Eq. \ref{eq:tauBar}), there exists a maximum grain size that still has efficient internal alignment, $a_{\max, aJ}$. Thus, the largest size of an aligned binary within the dust aggregate is described by $a_{\max,aJ}$, which depends on dust magnetic properties, grain rotation rate, and the gas density of the environment when grain growth occurs. As shown in Figure \ref{fig:amax_align}, for DC of $n_{\rm H}\lesssim 10^{5}\cm^{-3}$, dust of ordinary paramagnetic material has $a_{\max, aJ}\sim 0.5\mum$ for thermal rotation ($J/J_{d}=1)$ and $2\mum$ for suprathermal rotation ($J/J_{d}=100$), but it can increase to $5-10\mum$ for superparamagnetic grains with $N_{\rm cl}\sim 10^{4}$ iron atoms per cluster. For the typical protostellar core with $n_{\H}\sim 10^{7}-10^{8}\cm^{-3}$, only grains of sizes $a<a_{\max,aJ}\sim 2\mum$ can have internal alignment. For the interior of PPDs of $n_{\H}\gtrsim 10^{10}\cm^{-3}$, the maximum size of internally aligned grains is only $a_{\max,aJ}\lesssim 0.7\mum$, as implied by Equation (\ref{eq:aBar}). For grains larger than $a_{\max,aJ}$, internal relaxation is not efficient to align the grain's shortest axis with the magnetic field, but it may induce the wrong alignment with the shortest axis perpendicular to the magnetic field. In this case, dust aggregates may contain the aligned grains with parallel long axes due to grain-grain collisions.  

Note that in the diffuse ISM, large grains of highly elongated shapes and structures are found to be disrupted by centrifugal stress due to grain fast rotation by RATs (\citealt{Hoang:2019da}; \citealt{2019ApJ...876...13H}; see \citealt{2020Galax...8...52H} for a review). This process removes highly elongated shapes and leaves behind the less extreme grain shapes. 

\subsection{Implications for polarization observations toward star-forming regions}
Our study implies that grain growth from aligned grains has the elongation increasing from the diffuse ISM to MCs because the denser region has a higher growth rate, and the elongation can be large of $\epsilon>3$ (see Figure \ref{fig:elongation}). Therefore, the polarization cross-section efficiency (or intrinsic polarization efficiency) which is determined by the grain elongation would increase toward denser environments as long as grains are still aligned. Assuming the perfect alignment of grains with the magnetic field, then the maximum level of thermal dust polarization would increase and can be larger than the maximum level of $19.8\%$ measured by {\it Planck} satellite (\citealt{2015A&A...576A.104P}). Interestingly, high-resolution polarimetric observations of protostellar disks with {\it ALMA} report the maximum polarization level of $30-40\%$ (\citealt{2019ApJ...879...25K}; see Table 1 in \citealt{t9t}), which reveal that dust grains in these environments are more elongated than interstellar dust and consistent with our present prediction.

Observations of dust polarization in far-infrared and submm wavelengths usually reveal the decrease of dust polarization with increasing the visual extinction or column gas density. One popular explanation is that grains become less elongated toward dense regions due to grain growth (e.g., \citealt{2017A&A...597A..74J}). Such an explanation is expected for grain growth by gas accretion on randomly oriented grains because of the isotropic Brownian motion of thermal gas. However, our present study suggests that anisotropic gas accretion to aligned grains makes the grain more elongated. Therefore, the origin of the polarization hole toward dense starless clouds is most likely caused by the loss of grain alignment due to attenuation of radiation field and enhanced gas damping, as shown in \cite{Hoang.2021}, or magnetic field tangling. 


\subsection{Coagulation of aligned grains in PPDs and implications for cometary dust}
Comets form out of dust, ice, and gas beyond the snowline in PPDs, and thus, cometary dust aggregates contain valuable information about the coagulation process. However, where the growth from submicron sized grains to micron-sized dust aggregates starts, in PPDs or early phase of MCs, is uncertain.

Suppose that grain coagulation to micron-sized begins in MCs. In this case, it involves grains aligned with the magnetic field, and thus, produces an elongated binary comprising two aligned grains parallel short axes embedded within a dust aggregate. The maximum size of aligned grains,  $a_{\max,aJ}$, could reach several microns (see Figure \ref{fig:amax_align}). This fundamental structure represents the earliest phase of grain growth where the environment is still not dense enough and grains can still be aligned. The elongated binary with axially aligned grains would be imprint in cometary dust particles. Moreover, the elongation of dust grains resulting from gas accretion and coagulation should increase with the grain size. Since aligned grains require paramagnetic and superparamagnetic inclusions, the aligned binary and elongated grains would contain a high level of iron inclusions. 

Suppose that grain growth mainly occurs in the disk midplane of PPDs. In this case, only submicron grains of $a<a_{\max,aJ}\lesssim 0.7\mum$ can have internal alignment (see Eq. \ref{eq:aBar}), while larger grains with wrong internal alignment will lead to the dust aggregate made of aligned grains with parallel long axes. The orientation of the alignment axis in dust aggregates thus reveals unique information about where dust growth occurs.

Although radiation is weak, grains in PPDs may be spun-up by mechanical torques (\citealt{2007ApJ...669L..77L}; \citealt{2018ApJ...852..129H}; \citealt{kk7}) and rotates with high angular velocities. Thus, inelastic relaxation may be efficient to induce internal alignment for mm-sized grains (private communication, B.T. Draine). However, METs can both spinup and spindown the grain.

Carbonaceous grains are not likely to be aligned with the magnetic field due to their diamagnetic property (\citealt{2016ApJ...831..159H}). Although carbonaceous grains can still drift across the magnetic field as silicate grains, due to their random orientation, their shapes resulting from gas accretion is radically different from silicate grains. If stardust is shattered upon injected into the ISM (e.g., by shocks) and re-grows by gas accretion, then carbonaceous grains would have smaller elongation than silicate grains if these two populations are separate. As a result, the fundamental carbonaceous grains imprint in cometary dust would be less elongated than silicates. Carbonaceous grains would produce dust aggregate consisting of elongated grains with random axes. The identification of such a random orientation of elongated carbonaceous grains within dust aggregates would be a direct evidence for non-alignment of carbonaceous dust.

Observations of scattered light from cometary dust reveal circular polarization (CP; \citealt{Kolokolova:2016cd}). The alignment of cometary dust  grains is suggested as a potential mechanism to produce CP (\citealt{2014MNRAS.438..680H}). Scattering of sunlight by dust aggregates containing aligned grains would induce circular polarization, in analogy as aligned grains in the gas. If it is the case, CP can be produced as long as dust aggregates are lifted from the nucleus. 

\begin{figure}
\includegraphics[width=0.5\textwidth]{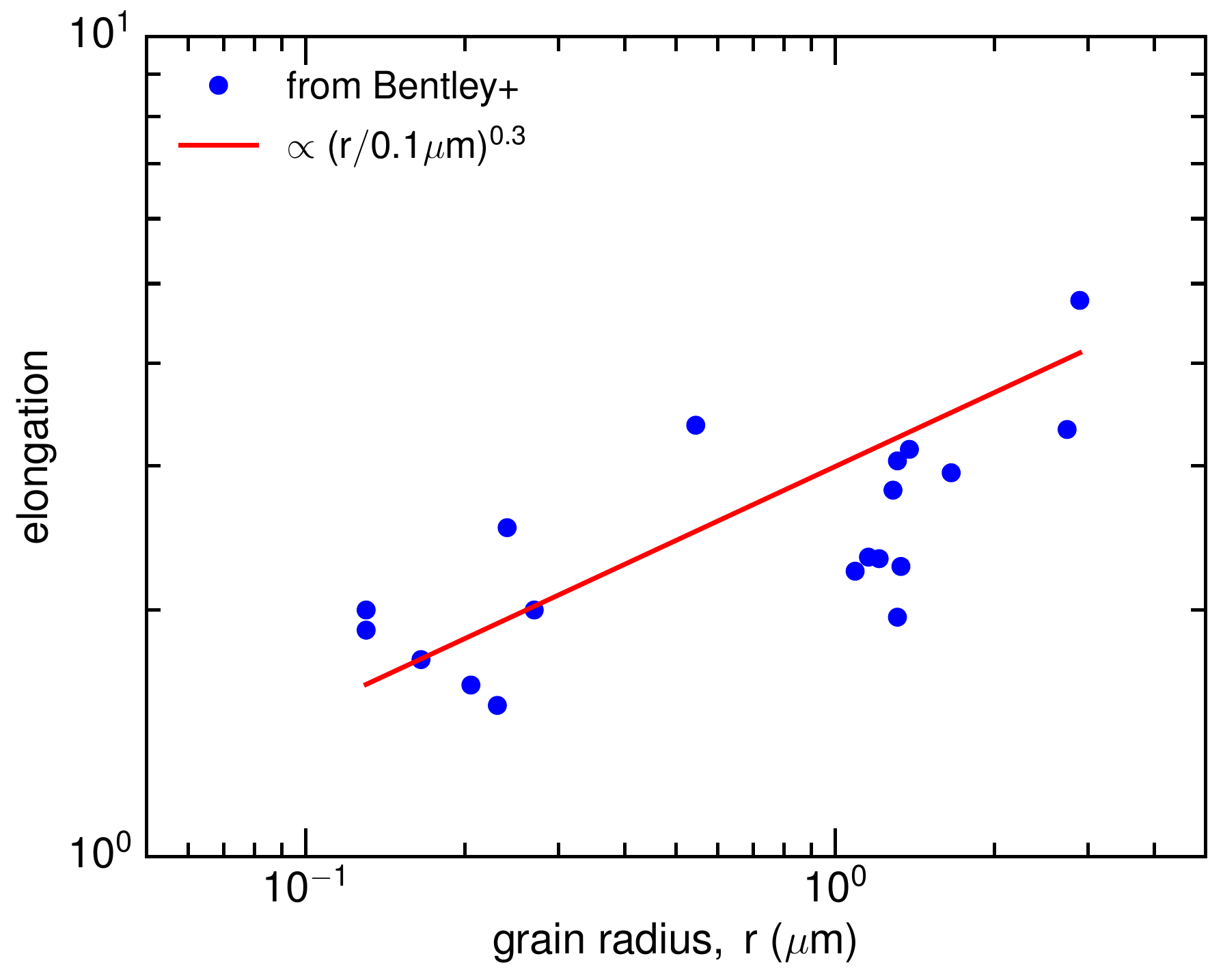}
\caption{Elongation versus the radius of dust grains locked in dust aggregates from 67P/Churyumov-Gerasimenko comet from \cite{2016Natur.537...73B}. The solid line denotes the slope of $0.3$. The grain elongation tends to increase with the grain radius, which supports the idea of grain growth from aligned grains.}
\label{fig:elongation}
\end{figure}

Finally, the {\it Rosetta} mission analyzed dust from 67P/Churyumov-Gerasimenko and found that cometary dust particles are aggregates of smaller, elongated grains of submicron sizes (\citealt{2016Natur.537...73B}). Figure \ref{fig:elongation} shows the variation of the fine grain elongation with the radius of dust grains within dust aggregates using data from \cite{2016Natur.537...73B}. The observed elongation appears to increase with its radius, which can be described by a slope of $0.3$ (solid line). This result reveals that grain growth is not isotropic, as expected from Brownian motion or randomly-oriented grains. Nevertheless, it is consistent with grain growth from gas accretion and grain coagulation involving aligned grains. Further studies on the fundamental aligned structures and elongation of aligned grains within cometary and interplanetary dust aggregates are essential to test this scenario of grain growth.

\section{Summary}\label{sec:summary}
We study the effect of grain alignment with the interstellar magnetic field on grain growth due to gas accretion and grain-grain collisions. The main results are summarized as follows.

\begin{enumerate}

\item Charged grains in the ISM can be accelerated by gyroresonance by MHD turbulence, ambipolar diffusion, or shocks, and move in the direction perpendicular to the magnetic field.

\item In dense MCs of $n_{\H}>10^{3}\cm^{-3}$, paramagnetic grains can have efficient internal alignment by Barnett relaxation with the shortest axis parallel to the magnetic field for grain sizes of $a\lesssim a_{\max,aJ}\sim 0.5-2\mum$ for the grain angular momentum of $J/J_{d}\sim 1-100$. The inclusion of iron clusters can shift the range of internal alignment grains to $a_{\max,aJ}\sim 10-50\mum$. 

\item Grain alignment by RATs for grains of $a<a_{\max,aJ}$ occurs with the shortest axis parallel to the magnetic field, so the long axis is thus parallel to the direction of grain motion. The grain alignment with the magnetic field occurs on a timescale shorter than grain growth time by gas accretion and coagulation, therefore, grain alignment should be considered for grain growth in MCs.

\item Fast rotation of the grain along the shortest axis (also magnetic field) leads to the oblate spheroidal shape for large grains that are grown by gas accretion to aligned elongated grains. The grain elongation by gas accretion increases over time and can exceed $\sim 3$ if the grain moves supersonically with respect to the metals.

\item Coagulation from aligned grains first produces a binary of aligned grains with parallel short axes and then to composite particles of oblate spheroid shapes due to the fast rotation around the magnetic field. The dust aggregates would contain a number of elongated binary comprising two aligned grains with parallel short axes.

\item Elongation of grains formed form coagulation of aligned grains should be larger than that formed from non-aligned grains. This is opposite from grain growth in the case of randomly oriented grains that makes the grain less elongated because the newly formed outer layer is isotropic. Due to their non-alignment with the magnetic field, carbonaceous grains would have lower elongation than silicate grains or grains with iron inclusions.

\item The orientation of elongated grains within a dust aggregate of cometary dust would provide information about where grain growth begins. Parallel short axes of aligned grains imply that dust aggregates are formed in MCs or the surface layer of PPDs where grains are aligned. A random orientation of elongated grains implies that grain growth begins in very dense, shielded regions where grains are not aligned due to high density and lack of radiation field.

\item Grains within dust aggregates in 67P/Churyumov-Gerasimenko studied by {\it Rosetta} have the elongation increasing with the grain radius, which implies that such dust aggregates might form by coagulation of aligned grains.

\end{enumerate}

\acknowledgments
We thank the referee for a constructive report and Bruce T. Draine, Ludmilla Kolokolova and Hiroshi Kamura for various comments. T.H. acknowledges the support by the National Research Foundation of Korea (NRF) grants funded by the Korea government (MSIT) through the Mid-career Research Program (2019R1A2C1087045).

\appendix
\section{Flux of gas accretion onto a drifting aligned grain}\label{sec:apdx}

Let $\hat{x}\hat{y}\hat{z}$ define the coordinate system of the lab frame with $\hat{z}\| {\bf B}$. We assume that an aligned dust grain drifts across the magnetic field with the drift velocity ${\bf v}_{d}=v_{d}\hat{x}$.  We consider the collision of the grain with a gas specie $i$ with mass $m_{i}$ and number density $n_{i}$. Let $s_{d}=v_{d}/v_{T,i}$ with $v_{T,i}=(2kT_{\rm gas}/m_{i})^{1/2}$ is the thermal velocity of the gas specie $i$.
 
The velocity distribution of the gas species in the lab frame is Maxwellian
\bea
f(v_{x},v_{y},v_{z})d{\bf v}=f_{x}f_{y}f_{z}dv_{x}dv_{y}dv_{z}=Z^{3}\exp\left(\frac{-m_{i}v^{2}}{2kT_{\rm gas}}\right)dv_{x}dv_{y}dv_{z}
\ena
where $Z=(m_{i}/2\pi kT_{\rm gas})^{1/2}$, and $\int_{-\infty}^{\infty} f({\bf v})dv_{x}dv_{y}dv_{z}=1$.

In the grain's reference frame, the gas velocity is $v'_{x}=v_{x}-v_{d},v'_{y}=v_{y},v'_{z}=v_{z}$.

The number flux of the gas species colliding with the upper surface of the grain along the z-direction is given by
\bea
F_{z}&=&\int_{-\infty}^{\infty} f_{x}dv_{x}\int_{-\infty}^{\infty}f_{y}dv_{y}\int_{0}^{\infty}n_{i}v_{z}f_{z}dv_{z}\nonumber\\
&=&n_{i}\int_{0}^{\infty}Zv_{z}e^{-m_{i}v_{z}^{2}/2kT_{\rm gas}}dv_{z}=Zn_{i}(2kT_{\rm gas}/m_{i})\int_{0}^{\infty}Ze^{-x}dx/2=\frac{1}{2}n_{x}\left(\frac{m_{i}}{2\pi kT_{\rm gas}}\right)^{1/2}\frac{2kT}{m_{i}}=\frac{1}{2}n_{i}\left(\frac{2kT_{\rm gas}}{\pi m_{x}}\right)^{1/2}\nonumber\\
&=&\frac{1}{4}n_{i}\langle v\rangle_{i}
\ena 
where $x=v_{z}^{2}/(2kT_{\rm gas}/m_{i})$ and the lower limit is taken to be zero in order for the gas to hit the grain surface, and $\langle v\rangle_{i}=(8kT_{\rm gas}/\pi m_{i})^{1/2}$ is the mean speed defined by $\langle v\rangle_{i}=\int_{0}^{\infty} v\times 4\pi v^{2}Z^{3}e^{-m_{i}v^{2}/2kT_{\rm gas}}dv$.

Similarly, the collision rate by the gas onto one upper surface along the y-direction is given by
\bea
F_{y}=\int_{-\infty}^{\infty} f_{x}dv_{x}\int_{-\infty}^{\infty}f_{z}dv_{z}\int_{0}^{\infty}n_{x}f_{y}dv_{y}
=\frac{1}{4}n_{i}\langle v\rangle_{i}.
\ena 

Due to the grain motion, only the gas atom with velocity $v'_{x}=v_{x}-v_{d}<0$ can collide with the grain. The collision rate to the front surface along the x-axis is
\bea
F_{x}=\int_{-\infty}^{\infty}Z\exp\left(-\frac{m_{i}v_{y}^{2}}{2kT_{\rm gas}}\right)dv_{y}\int_{-\infty}^{\infty}Z\exp\left(-\frac{m_{i}v_{z}^{2}}{2kT_{\rm gas}}\right)dv_{z}\int_{-\infty}^{v_{d}} dv_{x}n_{i}[-(v_{x}-v_{d})]Z\exp\left(-\frac{m_{i}v_{x}^{2}}{2kT_{\rm gas}}\right),
\ena
where the minus size before $(v_{x}-v_{d})$ accounts for the fact that the rate is positive. Performing the integral, one obtains
\bea
F_{x}&=&-\int_{-\infty}^{v_{d}} dv_{x}n_{i}(v_{x}-v_{d})Z\exp\left(-\frac{m_{i}v_{x}^{2}}{2kT_{\rm gas}}\right)=n_{i}\pi ac\left(\frac{1}{4}\langle v\rangle_{i} e^{-m_{i}v_{d}^{2}/2kT_{\rm gas}} + \frac{v_{d}}{2}[1+\erf(v_{d}(m_{i}/2kT_{\rm gas})^{1/2}]\right) \nonumber\\
&=&n_{i} \left(\frac{1}{4}\langle v\rangle_{i} e^{-s_{d}^{2}} +\frac{v_{d}}{2}[1+\erf(s_{d})]\right),
\ena
where $\erf(\alpha)=\frac{2}{\sqrt{\pi}}\int_{0}^{\alpha}e^{-x^{2}}dx$. For $v_{d}=0$, one see that $F_{x}=\frac{1}{4}n_{i}\langle v\rangle_{i}$. For $v_{d}\gg v_{T}$, the first term becomes negligible and the second term goes to $n_{i}v_{d}$ because $\erf(v_{d}\rightarrow \infty)\rightarrow 1$.

\section{Derivation of Barnett relaxation time}\label{apdx:internal}
For the pedagogical purpose here, we provide the detailed derivation of Barnett relaxation for an oblate spheroidal grain.

In the grain's body frame, the tips of $\bJ$ and $\bOmega$ precess around $\ahat_{1}$ with the angular rate $\omega=(h-1)\Omega_{1}$ (e.g., \citealt{1979ApJ...231..404P}; \citealt{Hoang:2010jy}) (see Figure \ref{fig:torque-free}). The rotational energy of the grain is
\bea
E_{\rm rot}(J,\theta)=\sum_{i=1}^{i=3}\frac{I_{i}\Omega_{i}^{2}}{2}=\sum_{i=1}^{i=3}\frac{J_{i}^{2}}{2I_{i}}=\frac{J_{1}^{2}}{2I_{1}}+\frac{J_{\perp}^{2}}{2I_{2}}
=\frac{J^{2}}{2I_{1}}\left[1+(h-1)\sin^{2}\theta\right],
\ena
where $J_{\perp}^{2}=J_{2}^{2}+J_{3}^{2}=J^{2}\sin^{2}\theta$.

The basic idea of this dissipation process is that, the instantaneous magnetization is along ${\bf M}=\chi(0) \bOmega/\gamma_{e}= \chi(0){\bf H}_{\rm BE}$ with $H_{\rm BE}\equiv\bOmega/\gamma_{e}$ being the equivalent-Barnett field. In the body frame, this magnetization has a component perpendicular to $\ahat_{1}$, which is seen rotating with respect to the axis of maximum inertia at angular rate $\omega$ (see Figure \ref{fig:torque-free}. This rotating magnetization induces the dissipation of rotational energy at rate,
\bea
\frac{dE_{\rm Bar}}{dt}&=& H_{\rm BE,rot}^{2}V\chi_{2} \omega =\left(\frac{\Omega \sin\theta_{\Omega}}{\gamma_{e}}\right)^{2}VK(\omega)\omega^{2},\nonumber\\
&=&\frac{\Omega^{2}\sin^{2}\theta_{\Omega}}{\gamma_{e}^{2}} VK(\omega)(h-1)^{2}\Omega_{1}^{2}
\ena 
where the rotating magnetization component $H_{\rm BE,rot}=H_{\rm BE}\sin\theta_{\Omega}$ with $\theta_{\Omega}$ the angle between $\bOmega$ and $\ahat_{1}$, $\chi_{2}(\omega)$ is the imaginary part of the complex magnetic susceptibility, and $K(\omega)=\chi_{2}(\omega)/\omega$. 

Using the relationship $\Omega_{1}=\Omega\cos\theta_{\Omega}=\Omega_{0}\cos\theta$ and $\Omega_{\perp}=\Omega \sin\theta_{\Omega}=(J/I_{\perp})\sin\theta$, so that $\Omega \sin\theta_{\Omega}=(J/I_{\|}) h\sin\theta=\Omega_{0}h\sin\theta$, one obtain
\bea
\frac{dE_{\rm Bar}}{dt}=\frac{\Omega^{2}\sin^{2}\theta_{\Omega}}{\gamma_{e}^{2}} VK(\omega)(h-1)^{2}\Omega_{0}^{2}\cos^{2}\theta=\frac{\Omega_{0}^{4}h^{2}\sin^{2}\theta}{\gamma_{e}^{2}} VK(\omega)(h-1)^{2}\cos^{2}\theta=\frac{J^{4}}{I_{1}^{4}\gamma_{e}^{2}} VK(\omega)h^{2}(h-1)^{2}\sin^{2}\theta\cos^{2}\theta.\label{eq:dEbar_dt}
\ena

The Barnett dissipation induces the reduction of the rotational energy as
\bea
\frac{dE_{\rm rot}}{dt}=\frac{J^{2}}{I_{1}}(h-1)\sin\theta\cos\theta \dot{\theta},\label{eq:dErot_dt}
\ena

Using $dE_{\rm rot}/dt=-dE_{\rm Bar}/dt$ and Equations (\ref{eq:dEbar_dt}) and (\ref{eq:dErot_dt}) one obtains
\bea
\frac{d\theta}{dt}=-\frac{J^{2}}{I_{1}^{3}\gamma_{e}^{2}} VK(\omega)h^{2}(h-1)\sin\theta\cos\theta
\ena
which corresponds to
\bea
\frac{d\tan\theta}{\tan\theta} = -\frac{VK(\omega)h^{2}(h-1)J^{2}}{I_{1}^{3}\gamma_{e}^{2}}  dt=-\frac{dt}{\tau_{\rm Bar}}\rightarrow \tan\theta=Ce^{-t/\tau_{\rm BR}}
\ena
where $\tau_{\rm BR}$ is characteristic timescale for Barnett relaxation
\bea
\tau_{\rm BR}&=& \frac{\gamma_{e}^{2}I_{\|}^{3}}{VK(\omega)h^{2}(h-1)J^{2}}.~\label{eq:tauBar_apdx}
\ena

\bibliography{ms.bbl}
\end{document}